\documentclass[VANCOUVER,STIX1COL]{WileyNJD-v2}
\usepackage[T1]{fontenc}
\usepackage{ifthen} % Conditional commands in LATEX documents
\usepackage{xspace}%to cope with special spacing after figure and table references
\usepackage{color}% To use colors for \TOFIX commands;  \definecolor command for defining your own colors
\usepackage{xcolor}% Add more colours
\usepackage{paralist}

\usepackage{algorithm}
\usepackage{algpseudocode}
\usepackage{makecell}
\algnewcommand{\LineComment}[1]{\State \(\triangleright\) #1}

\usepackage{rotating}
\usepackage[export]{adjustbox}

\usepackage{tabulary}

\newcommand{\excluded}[1]{#1}
\renewcommand{\excluded}[1]{} %Comment out this whole line to surpress exclusion.
\newboolean{fancylst}
\setboolean{fancylst}{true}
\ifthenelse{\boolean{fancylst}} {
	\usepackage[most]{tcolorbox}

	\newtcblisting[auto counter]{fancylisting}[2][]{
		sharp corners,
		top=-2mm,
		bottom=-2mm,
	    fonttitle=\bfseries,
	    colframe=gray,
	    listing only,
	    enhanced, 
	    listing options={
	    	numbers=left,
	       	numberstyle=\fontsize{8.1}{10}\ttfamily,
	      	basicstyle=\fontsize{8.1}{10}\ttfamily ,
	    	breaklines= true,
	     	showspaces= false,
	     	numbersep=2.5mm,
	     	language=c++,
	     	tabsize=4,
	     	escapeinside={(*@}{@*)}},
		overlay={%
	        \begin{tcbclipinterior}
	            \fill[gray!25] (frame.south west) rectangle ([xshift=4mm]frame.north west);
	        \end{tcbclipinterior}},
	    title=Listing \thetcbcounter: #2,
	    #1} %this is the .label
}{

\lstnewenvironment{fancylisting}[2][]{%
\lstset{
	caption=#2,
	#1, %this is the .label
    frameround=ffff,
    frame=single,
	language=c++,
	numbers=left,
    stepnumber=1,
    showstringspaces=false,
    tabsize=4,
    breaklines=true,
    breakatwhitespace=false,
    escapeinside={(*@}{@*)}}
	}%
	{}
}

\newboolean{anonymous}
\setboolean{anonymous}{false}
\newboolean{includecomments}
\setboolean{includecomments}{true}
\newcommand{\Clang}{\textsf{Clang}\xspace}
\newcommand{\LLVM}{\textsf{LLVM}\xspace}
\newcommand{\LLVMIR}{\textsf{LLVM IR}\xspace}

\newcommand*{\RQOne} {Isn't this a splendid research question here ?}
\newcommand*{\RQTwo} {And isn't this one even better ?}
\newcommand*{\RQThree} {This one tops it all, doesn't it ?}
\newcommand*{\RQFour} {Except, isn't this one the best ?}

\ifthenelse{\boolean{anonymous}} {
    \newcommand{\Dextool}{\textsf{Anonymised}\xspace}
}{
    \newcommand{\Dextool}{\textsf{Dextool mutate}\xspace}
}
\newcommand{\hybrid}{\text{reachable schemata}\xspace}

\newcommand{\myparagraph}{\par\null\noindent\textit}

\newcommand{\secref}[1]{Section~\ref{#1}\xspace}
\newcommand{\figref}[1]{Figure~\ref{#1}\xspace} 
\newcommand{\tabref}[1]{Table~\ref{#1}\xspace}
\newcommand{\algoref}[1]{Algorithm~\ref{#1}\xspace}
\newcommand{\formularefpage}[1]{Formula~\ref{#1}~--~p.~\pageref{#1}\xspace}

\usepackage{fancybox}%Pro?vides vari?ants of \fbox

\newcommand{\hypobox}[1]{
	\begin{center}%
        \noindent\thicklines\setlength{\fboxsep}{2pt}%
        \cornersize{0.1}
        \ovalbox{\begin{minipage}{\textwidth}%
		#1
		\end{minipage}}
	\end{center}}

\ifthenelse{\boolean{includecomments}} {
    \newcommand{\nb}[3]{
    	{\colorbox{#2}{\bfseries\sffamily\scriptsize\textcolor{white}{#1}}}
    	{\textcolor{#2}{\sf$\blacktriangleright$\textit{#3}$\blacktriangleleft$}}}
} {
    \newcommand{\nb}[3]{}
}

\newcommand{\paperTitle}{Mutation Testing Optimisations using the Clang Front-end}
\newcommand{\submitTo}{Software: Testing, Verification and Reliability}

\hypersetup{
    colorlinks,%
    citecolor=black,%
    filecolor=black,%
    linkcolor=black,%
    urlcolor=gray,
    linktocpage=true,
    bookmarks=true,
    bookmarksopen=true,
    pdfpagemode=UseOutlines,
    pdftitle={\paperTitle},    % title
    pdfauthor={Anonymous},     % author
    pdfsubject={\submitTo },   % subject of the document
    pdfcreator={PDF LaTex},   % creator of the document
}

\articletype{Article Type}%

\received{26 April 2016}
\revised{6 June 2016}
\accepted{6 June 2016}

\begin{document}

\title{\paperTitle}

\author[1]{Sten Vercammen*}

\author[1,2]{Serge Demeyer}

\author[3,4]{Markus Borg}

\author[5]{Niklas Pettersson}

\author[4]{G\"{o}rel Hedin}

\authormark{STEN VERCAMMEN \textsc{et al}}

\address[1]{\orgdiv{Dept. of Informatics}, \orgname{University of Antwerp}, \orgaddress{\state{Antwerp}, \country{Belgium}}}

\address[2]{%\orgdiv{Org Division},
\orgname{Flanders Make}, \orgaddress{\state{Antwerp}, \country{Belgium}}}

\address[3]{\orgdiv{Digital Systems}, \orgname{RISE Research Institute of Sweden}, \orgaddress{\state{Lund}, \country{Sweden}}}

\address[4]{\orgdiv{Dept. of Computer Science}, \orgname{Lund University}, \orgaddress{\state{Lund}, \country{Sweden}}}

\address[5]{\orgdiv{Saab Aeronautics}, \orgname{Saab AB}, \orgaddress{\state{Linköping}, \country{Sweden}}}

\corres{*Sten Vercammen, \email{sten.vercammen@uantwerpen.be}}

\presentaddress{Universiteit Antwerpen -- Department of Computer Science. Middelheimlaan 1, BE-2020 ANTWERPEN}

\abstract[Summary]{Mutation testing is the state-of-the-art technique for assessing the fault detection capacity of a test suite.
Unfortunately, a full mutation analysis is often prohibitively expensive.
The CppCheck project for instance, demands a build time of  5.8 minutes and a test execution time of 17 seconds on our desktop computer.
An unoptimised mutation analysis, for 55,000 generated mutants took 11.8 days in total, of which 4.3 days is spent on (re)compiling the project.
In this paper we present a feasibility study, investigating how a number of optimisation strategies can be implemented based on the Clang front-end.
These optimisation strategies allow to eliminate the compilation and execution overhead in order to support efficient mutation testing for the C language family.
We provide a proof-of-concept tool that achieves a speedup of between 2x and 30x.
We make a detailed analysis of the speedup induced by the optimisations, elaborate on the lessons learned and point out avenues for further improvements.
}


\jnlcitation{\cname{%
\author{Sten Vercammen}, 
\author{Serge Demeyer}, 
\author{Markus Borg},
\author{Niklas Pettersson},
\author{G\"{o}rel Hedin}} (\cyear{2022}), 
\ctitle{Mutation Testing Optimisations using the Clang Front-end},
\cjournal{Journal of Software: Testing, Verification and Reliability}, \cvol{2022;00:xx--xx to appear}.}

\maketitle

% !TEX root = Vercammen2022STVR.tex

\section{Introduction}
\label{sec::Intro}

DevOps (Software Development and IT Operations) is defined by Bass \etal{} as ``\textit{a set of practices intended to reduce the time between committing a change to a system and the change being placed into normal production, while ensuring high quality}''~\cite{Bass2015DevOps}.
This allows for frequent releases to rapidly respond to customer needs.
Tesla, for example, uploads new software in its cars once every month~\cite{TeslaUpdates};
Amazon pushes new updates to production on average every 11.6 seconds~\cite{Jenkins2011}.
The enabling factor for the DevOps approach is a series of automated tests which serve as quality gates, safeguarding against regression faults.

The growing reliance on automated software tests raises a fundamental question: How trustworthy are these automated tests?
Today, mutation testing is acknowledged within academic circles as the most promising technique for assessing the \textit{fault-detection capability} of a test suite~\cite{jia2011analysis, Papadakis2019}.
The technique deliberately injects faults (called mutants) into the production code and counts how many of them are caught by the test suite.
The more mutants the test suite can detect, the higher its fault-detection capability is -- referred to as the \textit{mutation coverage} or \emph{mutation score}.

Case studies with safety-critical systems demonstrate that mutation testing could be effective where traditional structural coverage analysis and code inspections have failed~\cite{Baker2013mutationtestingsafetrycritical, Ramler2017mutationtestingsafetrycritical}.
Google on the other hand reports that mutation testing provides insight into poorly tested parts of the system, but --more importantly-- also reveals design problems with components that are difficult to test, hence must be refactored~\cite{Petrovic2018mutationtestingatgoogle}.
In a similar vein, a blog post from a software engineer at NFluent, comments on integrating Stryker (a mutation tool for .Net programs) in their development pipeline~\cite{Dupuydauby2019Stryker}.
There as well, mutation testing revealed weaknesses in the test suite, but also illustrated that refactoring allowed for simpler test cases which subsequently increased the mutation score.

Despite the apparent potential, mutation testing is difficult to adopt in industrial settings.
One of the reasons is because the technique ---in its basic form--- requires a tremendous amount of computing power.
Without optimisations, the entire code base must be compiled and tested separately for each injected mutant~\cite{jia2011analysis}.
During one of our experiments with an industrial code base, we witnessed 48 hours of mutation testing time on a test suite comprising 272 unit tests and 5,256 lines of test code for a system under test comprising 48,873 lines of production code~\cite{vercammen2018speeding}.
Hence for medium to large test suites, mutation testing without optimisations becomes prohibitively expensive.

As a consequence, the last decades has devoted a lot of research to optimise the mutation testing process~\cite{Polo2010SurveyOptimisation, Papadakis2019}.
One stream of work focuses on \textit{parallelisation}, either on dedicated hardware~\cite{offutt1992mutation} or in the cloud~\cite{vercammen2018speeding}.
Another stream of work focuses on \textit{incremental approaches}, limiting the mutation analysis to what has been changed since the previous run~\cite{Wei2021MUdelta}.
A third stream of work (and the inspiration for what is presented in this paper) focuses on techniques based on program analysis, for example \textit{mutant schemata}~\cite{demillo1991compiler} and \textit{split-stream mutation testing}~\cite{king1991fortran}.
The former injects all mutants simultaneously, analysing the abstract syntax tree to ensure that the mutated version actually compiles.
The latter forks the test execution from the mutation point via a combination of static and dynamic program analysis.

Mutation testing shines in systems with high statement coverage because uncaught mutants reveal weaknesses in code which is supposedly covered by tests.
Safety-critical systems ---where safety standards dictate high statement coverage--- are therefore a prime candidate for validating optimisation strategies.
In safety-critical software, C and C++ dominate the technology stack~\cite{laplante2017software}.
Yet in the mutation testing community, the C language family is somehow neglected: a systematic literature review on mutation testing from 2019 reports that less than 25\% of the primary studies target source code from the C language family~\cite{papadakis2019mutation}.
This opens up opportunities as the C language family is a mature technology with considerable tool support available.
In particular, the compiler front-end \Clang that operates in tandem with the \LLVM compiler back-end~\cite{CLANG13}.

This paper presents a feasibility study, investigating to which extent the \Clang front-end and its state-of-the-art program analysis facilities allow to implement existing strategies for mutation optimisation within the C language family.
We present a proof-of-concept optimisation tool, featuring a series of representative optimisation strategies.
These optimisation strategies are:
\begin{compactenum}
\item \textit{exclude invalid mutants}: avoid compilation overhead from mutants that would cause downstream compilation errors.
\item \textit{exclude unreachable mutants}: avoid execution overhead from mutants that are not reached by the test suite.
\item \textit{mutant schemata}: where all mutants get injected simultaneously and the project is only compiled once~\cite{demillo1991compiler}.
At test execution time, the appropriate mutant is selected via a boolean flag.
\item \textit{reachable mutant schemata}: an extension of mutant schemata which reduces the test suite to only those test cases which reach the mutant.
\item \textit{split-stream mutation testing}: where tests are executed from the mutation point itself, by forking the process, essentially avoiding redundant executions~\cite{king1991fortran}.
\end{compactenum}

We validate the proof-of-concept tool on four open-source C++ libraries and one industrial component.
These cases cover a wide diversity in size, C++ language features used, compilation times, and test execution times.
Hence, they may serve as a representative benchmark to validate mutation optimisation strategies.
To illustrate the potential of \Clang-based implementation of mutation optimisation strategies, we report the overhead induced and the speedup achieved, both in absolute as well as relative terms.
Furthermore, we derive a series of lessons learned on the benefits and impediments of implementing mutation testing optimisations using the \Clang compiler front-end.

The rest of the paper is structured as follows.
In \secref{sec::Background}, we elaborate on the concept of mutation testing and list related work.
In \secref{sec::proofOfConcept}, we describe the design of our proof-of-concept tool.
In \secref{sec::Exp_setup}, we explain how we obtained the speedups induced by the different optimisation strategies.
This naturally leads to \secref{sec::Results_Discussion} where we discuss the results and \secref{sec::LesonsLearned} where we derive the lessons learned.
As with any empirical research validating proof-of-concept tools, our study is subject to various threats to validity and limitations which are listed in \secref{sec::Threats}.
Finally we draw conclusions in \secref{sec::Conclusion}.

%%%%%%%%%%%%%%%%%%%%%%%%%%%%%%%%%%%

\section{Background and Related Work}
\label{sec::Background}
In this section, we elaborate on the concept of mutation testing, the different optimisation strategies as available in the related work and discuss existing work mutation optimisation based on the \Clang front-end.

\subsection{Mutation Testing Terminology}
For effective testing, software teams need strong tests which maximise the likelihood of exposing faults~\cite{myers197977ie}.
Traditionally, the strength of a test suite is assessed using code coverage, revealing which statements are poorly tested.
However, code coverage has been shown to be a poor indicator of test effectiveness~\cite{cai2005effect, inozemtseva2014coverage}.
Stronger coverage criteria, like full MC/DC coverage (Modified Condition/Decision Coverage, a coverage criterion often mandated by functional safety standards that target critical software systems, e.g., ISO~26262 and DO-178C) still does not guarantee the absence of faults~\cite{Gay2015riskcoverage, Kandl2015}.
Today, mutation testing (sometimes also named \emph{mutation analysis}, the terms are used interchangeably) is the state-of-the-art technique for assessing the \textit{fault-detection capacity} of a test suite~\cite{jia2011analysis, Papadakis2019}.

\textit{Killed / survived mutants.}
Mutation testing deliberately injects faults (called mutants) into the production code and counts how many of them are caught by the test suite.
A mutant caught by the test suite, i.e. at least one test case fails on the mutant, is said to be \emph{killed}.
When all tests pass, the mutant is said to \emph{survive}.

\textit{Mutation Coverage.}
A strong test suite should have as few surviving mutants as possible.
This is expressed in a score known as the mutation coverage: the number of mutants killed divided by the total number of mutants injected.
A high mutation coverage implies that most mutants get killed; 100\% is a perfect score as the tests can reveal all small deviations.
Mutation coverage is sometimes referred to as mutation score.

\begin{table}[!htbp]
   \centering
   \begin{tabular}{ l p{5cm} p{10cm}}
   	\hline
	\textbf{Code}    & \textbf{Short} & \textbf{Decription}\\
	\hline
	ROR & Relational Operator Replacement & Replace a single operator with another operator.
The relational operators are $<$,$<=$,$>$,$>=$,$==$,$!=$ \\
	AOR & Arithmetic Operator Replacement & Replace a single arithmetic operator with another operator.
The operators are: $+, -, *, /, \%$ \\
	LCR & Logical Connector Replacement & Replace a logical connector with the inverse.
The logical connectors are: $||$, $\&\&$, $|$, $\&$ \\
	UOI & Unary Operator Insertion & Insert a single unary operator in expressions.
Example unary operators are: increment ($++$), decrement ($--$), address ($\&$), indirection ($*$), positive ($+$), negative ($-$), \ldots\\
	SDL & Statement Deletion &  Selective deletion of code, including removing a specific function call, or replacing a method body by \texttt{void}\\
	AMC & Access Modifier Change & Changes the access level for instance variables and methods to other access levels.
Access levels are \texttt{private}, \texttt{protected}, \texttt{public}\\
	ISI & Super Keyword Insertion &  Inserts the \texttt{super} keyword so that a reference to the variable or the method goes to the overridden instance variable or method\\
	ICR & Integer Constant Replacement & Replaces every constant \textit{c} with a value from the set $\{-1, 0, 1, -c, c-1, c+1\}\backslash\{c\}$ \\
	\hline
   \end{tabular}
   \caption{Overview of commonly used mutation operators for C and C++.}
   \label{tab:mutationoperators}
\end{table}

\textit{Mutation Operators.}
Mutation testing mutates the program under test by artificially injecting a fault based on a mutation operator.
A mutation operator is a source code transformation which introduces a change into the program under test.
Typical examples are replacing a conditional operator (e.g., $>=$ into $<$) or an arithmetic operator (e.g., $+$ into $-$).
The first set of mutation operators was reported in King \etal~\cite{King1991}.
Later, special purpose mutation operators have been proposed to exercise errors related to specific language constructs, such as Java null-type errors~\cite{Parsai2019} or C++11/14 lambda expressions and move semantics~\cite{Parsai2018}.
There are more than 100 mutation operators reported in the academic literature and there is no consensus of which ones are best for a specific language and code base.
Therefore mutation testing tools feature a diverse set of mutation operators which can be configured when performing the mutation analysis.
\tabref{tab:mutationoperators} lists commonly used mutation operators for C and C++ which we will refer to later in this paper.

\textit{Invalid Mutants.}
Mutation operators introduce syntactic changes, hence may cause compilation errors in the process.
If we apply the arithmetic mutation operator (AOR) to e.g. ``a * b'', then we get four mutants as shown in Listing \ref{lst:Mutation Example}.
However, the modulo operator (`\%') will give an ``\texttt{invalid operands to binary expression}'' error, as the modulo operator is not defined for floating point data types.
The mutant can thus not be compiled and is considered invalid.

Another frequently occurring invalid mutant occurs when changing the `$+$' into a `$-$' which works for numbers but does not make sense when applied to the C++ string concatenation operator.
If the compiler cannot compile the mutant for any reason, the mutant is considered invalid and is not incorporated into the mutation coverage.
Preventing the generation of invalid mutants is one way to optimise the mutation testing process.

\begin{fancylisting}{Mutation Example,label=lst:Mutation Example}
float f(float a, float b) {
	return a * b;	// original code
}
	return a + b;	// mutant 1
	return a - b;	// mutant 2
	return a / b;	// mutant 3
	return a % b;	// mutant 4
          (*@ \textasciitilde\textasciitilde\textasciitilde \textasciicircum\textasciitilde\textasciitilde\textasciitilde \textcolor{red}{Invalid operands to binary expression} @*)
\end{fancylisting}

\textit{Unreachable Mutants.}
The Reach--Infect--Propagate--Reveal criterion (a.k.a. the RIPR model) provides a fine-grained framework to assess weaknesses in a test suite which are conveniently revealed by mutation testing~\cite{Li2016testoracle}.
It states that an effective test should first of all \textit{Reach} the fault, then \textit{Infect} the program state, after which it should \textit{Propagate} as an observable difference, and eventually \textit{Reveal} the presence of a fault.
When a mutant is injected in a statement that is never executed by the test suite, it can never be killed.
Therefore, one can optimise the mutation testing process by explicitly excluding unreachable mutants.
To increase its effectiveness, this should be done on the test case level instead of on the entire test suite.

\textit{Equivalent Mutants.}
Injected mutants can be syntactically different from the original software system, but semantically identical.
These mutants do not modify the meaning of the original program, and can therefore not be detected by the test suite. Such mutants are called equivalent mutants.
They yield false negatives and decrease the effectiveness of the mutation analysis.
Additionally, they waste developer time, as they need to be manually labelled as equivalent mutants because they show up as survived mutants, which can never be killed.
Consequently, a big challenge of mutation is handling (and/or eliminating) these equivalent mutants. An overview of techniques to overcome the equivalent mutant problem has been provided by Madeyski et al~\cite{madeyski2014overcoming}.
One of the frequently used techniques is called \textit{Trivial Compiler Equivalence (TCE)}: if the compiler emits the same low-level code (machine code) then the mutant is guaranteed to be equivalent.

\textit{Timed out Mutants.}
Some injected mutants cause the test to go into an infinite loop.
To prevent such infinite loops, mutants with an excessively long execution time need to be detected and stopped.
This is often done using a time out, which terminates the test after a predefined period of time.
A mutant which causes such a time-out is considered killed.

\subsection{Mutation Testing Optimisations}
\label{sec:optimisations}

To explain the time-consuming nature of the mutation testing process, \algoref{code:Mutation Testing} shows the essential steps of a mutation analysis without any optimisations.
The software system needs to build without errors and all software tests should succeed before mutation testing can even begin; this is called the \textit{pre}-phase.
Then, the two main phases are executed: (\textit{A}) the generate mutants phase and (\textit{B}) the execute mutants phase.
In phase \textit{A}, mutants are generated for all source files.
In phase \textit{B}, for each mutant, all tests are executed and the result ---whether or not it was killed--- is saved.
Finally, all the results are gathered and the final report is created in the \textit{post}-phase.

\begin{algorithm}[hbt]
\caption{Pseudocode Mutation Testing}
\label{code:Mutation Testing}
\begin{algorithmic}[1]
\Function{mutationTesting}{srcFolder $src$}
	\LineComment{Pre: verify build and if all tests succeed}
	\If {$\Call{initialBuildAndTests}{ } \not= \textbf{True}$}
		\State \Return
	\EndIf
	\State
	\LineComment{A: generate mutants}
	\State $mutants\gets []$
	\ForAll{srcFile $f \in src$}
		\State $fMutants\gets \Call{generateMutants}{$srcFile $f}$
		\State $mutants\gets mutants + fMutants$
	\EndFor
	\State
	\LineComment{B: execute mutants}
	\ForAll{mutant $m \in mutants$}
		\State $\Call{compileMutant}{$mutant $m$}
		\State $result\gets \Call{executeMutant}{$mutant $m$, testsuite $t}$
		\State \Call{storeResult}{$result$, mutant $m$, testsuite $t$}
	\EndFor
	\State
	\LineComment{Post: process results}
	\State $\Call{processResults}{ }$
\EndFunction
\end{algorithmic}
\end{algorithm}

A lot of research is devoted to optimising the mutation testing process, summarised under the principle --- \textit{do fewer}, \textit{do smarter}, and \textit{do faster}~\cite{offutt2001mutation}.
\begin{compactitem}
\item \textit{Do fewer} approaches minimise the execution time by reducing the total amount of mutants to execute.
Such an optimisation can be implemented by generating fewer mutants on line 10 in \algoref{code:Mutation Testing} or by selecting a subset of all mutants on line 15.
Incremental approaches\cite{Wei2021MUdelta}, limiting the mutation analysis to code changed in a commit are a particularly relevant example of a ``do fewer" approach.
A reduced set of mutants normally incurs an information loss compared to the full set of mutants, however, the effectiveness is often acceptable~\cite{jia2011analysis}.
Nevertheless, when the mutation testing tool provides sufficient guarantees to identify \textit{invalid mutants} or \textit{unreachable mutants}, excluding these is always an effective optimisation.

\item \textit{Do smarter} approaches attempt to minimise the execution time by exploiting the computer hardware (e.g. distributed architectures, vector processors, fast memory access).
The \texttt{for} loops in lines 9 and 15 in \algoref{code:Mutation Testing}) have few data dependencies, hence can be executed in parallel.
Parallel execution of mutants, either on dedicated hardware~\cite{offutt1992mutation} or in the cloud~\cite{vercammen2018speeding} is known to speed up the process by orders of magnitude.
\textit{Split-stream mutation testing} is the de facto representative for ``do smarter'' optimisations~\cite{king1991fortran}.
By retaining state information between test runs, split-stream mutation testing avoids the redundant execution of statements up until the mutation point.

\item \textit{Do faster} approaches attempt to minimise the execution time by reducing the execution cost for each mutant (cfr. line 17 in \algoref{code:Mutation Testing}).
By design, a mutated program is almost identical to the original program which can be exploited during the compilation step.
\textit{Mutant schemata}~\cite{demillo1991compiler} is the best know example.
All mutants get injected simultaneously (guarded by a global switch variable), hence the project is compiled only once in line 16.
During the mutant execution phase (in line 17) the global switch is used to select the actual mutant to execute.
The execution time of a mutant can also be reduced with \textit{test prioritisation} techniques.
By rearranging the test suite, the tests with the highest likelihood of failure will be executed first, reducing the test suite run time using early-failure~\cite{zhang2013faster}.

\item \textit{Hybrid approaches.}
Different approaches are often synergistic, where a combination of techniques becomes more than the sum of the parts.
Note that some approaches are orthogonal to one another hence are easy to combine.
Excluding unreachable mutants, for example, can be combined with any other optimisation.
Other approaches, however, may depend on each other.
Mutant schemata, for instance, requires that all invalid mutants are excluded because even a single invalid mutant will immediately invalidate the whole mutated program.
Measuring the speedup of a given optimisation strategy should take these synergies into account.
\end{compactitem}

\subsection{\LLVM \& \Clang Compiler Infrastructure}

\begin{quotation}
\textit{The \LLVM Project is a collection of modular and reusable compiler and toolchain technologies.
[...] capable of supporting both static and dynamic compilation of arbitrary programming languages.}~[\url{https://LLVM.org}]
\end{quotation}

\begin{figure*}[hbt]
	\centering
	\caption{LLVM Compilation}
	\label{fig:LLVM Compilation}
	\includegraphics[max size={0.75\textwidth}{\textheight}]{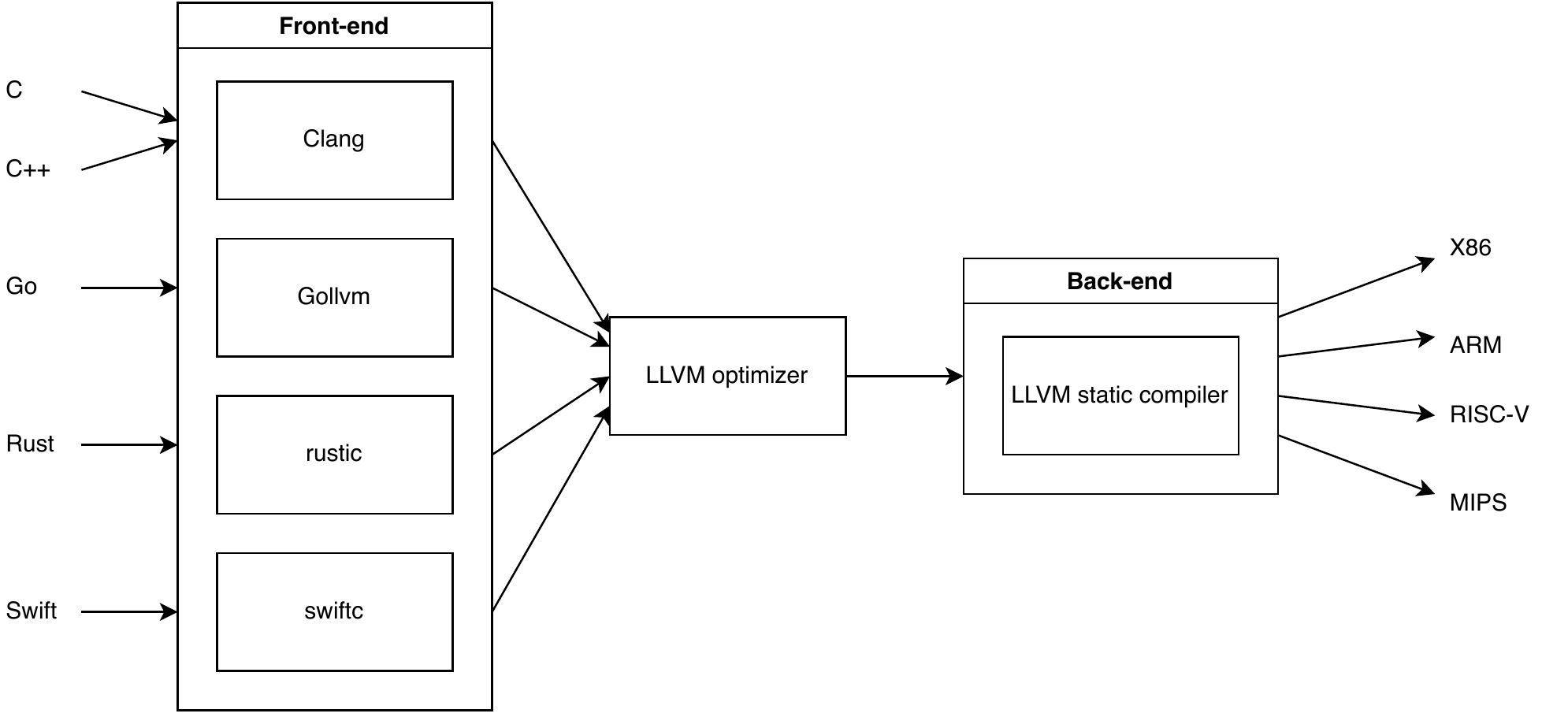}
\end{figure*}

LLVM is a collection of compilation tools designed around a low-level language-independent intermediate representation, the \LLVMIR.
The project includes frontends that translate source code to \LLVMIR, optimisers that rewrite the \LLVMIR to become faster, and backends that generate machine code from the \LLVMIR for different architectures.
We visualised this in \figref{fig:LLVM Compilation}.
\Clang is the most well-known frontend for \LLVM, and supports languages in the C family, like C, C++, and Objective-C, among others.
Internally, it represents programs as abstract syntax trees (ASTs).
\Clang includes a semantic analyser that does type-checking and other compile-time checks, before generating the IR.
Furthermore, \Clang contains a number of libraries based on the visitor pattern, allowing more analyses or transformations to be added to the frontend.

\LLVM and \Clang serve as the de facto standard for building static analysis tools for the C language family.
It should therefore come as no surprise that several C and C++ mutation tools exist that build upon these frameworks.
These tools mutate the program either at the AST level or at the \LLVMIR level.
Both approaches have advantages and disadvantages.
Doing the mutations at the \LLVMIR level has the advantage that they will work for any frontend, but the disadvantage that mutants injected in the \LLVMIR are difficult or even impossible to trace back to a source representation in the original code under test, which allows for the generation of many invalid mutants.

The AST, on the other hand, is close to the source code.
Mutating at this level provides good traceability, which is critically important for reporting back results to the developer in the \textit{post}-phase (lines 18 and 22 in \algoref{code:Mutation Testing}).
Another advantage of mutating at the AST level is that the frontend semantic analyser can be used to ensure that the mutated code is compile-time correct, effectively eliminating invalid mutants.

\subsection{Mutating on the \LLVMIR and AST level}
As the AST is close to the source code, the source code from Listing \ref{lst:Mutation Example} can be transformed into an AST without losing any information.
This can be seen in Listing \ref{lst:AST Example} where even the original line and column information of the statements are retained.
The AST includes all the semantic details in an easily accessible manner.
This, together with the frontend semantic analyser allows for more informed mutations, allowing the detection of invalid mutations such as the modulo operator of mutant 4.
Mutating the multiplication, i.e. the binary operator \verb+'+*\verb+'+ on line 6, is achieved by changing the \verb+'+*\verb+'+ to \verb+'++,-,/\verb+'+.
As the location information is preserved, the mutants are easily represented in the original source code.
Tools mutating the AST will do so by either iterating through the entire AST or by using AST matchers which provide a list of all candidate nodes based on a set configuration e.g. BinaryOperator.

\begin{fancylisting}{AST Example,label=lst:AST Example}
|-FunctionDecl 0x7fac9d87b710 <main.cpp:1:1, line:3:1> line:1:7 used f (*@\verb+'+@*)float (float, float)(*@\verb+'+@*)
| |-ParmVarDecl 0x7fac9d87b5b8 <col:9, col:15> col:15 used a (*@\verb+'+@*)float(*@\verb+'+@*)
| |-ParmVarDecl 0x7fac9d87b638 <col:18, col:24> col:24 used b (*@\verb+'+@*)float(*@\verb+'+@*)
| (*@\verb+`+@*)-CompoundStmt 0x7fac9d87b8a8 <col:27, line:3:1>
|   (*@\verb+`+@*)-ReturnStmt 0x7fac9d87b898 <line:2:5, col:16>
|     (*@\verb+`+@*)-BinaryOperator 0x7fac9d87b878 <col:12, col:16> (*@\verb+'+@*)float(*@\verb+'+@*) (*@\verb+'+@*)*(*@\verb+'+@*)
|       |-ImplicitCastExpr 0x7fac9d87b848 <col:12> (*@\verb+'+@*)float(*@\verb+'+@*) <LValueToRValue>
|       | (*@\verb+`+@*)-DeclRefExpr 0x7fac9d87b808 <col:12> (*@\verb+'+@*)float(*@\verb+'+@*) lvalue ParmVar 0x7fac9d87b5b8 (*@\verb+'+@*)a(*@\verb+'+@*) (*@\verb+'+@*)float(*@\verb+'+@*)
|       (*@\verb+`+@*)-ImplicitCastExpr 0x7fac9d87b860 <col:16> (*@\verb+'+@*)float(*@\verb+'+@*) <LValueToRValue>
|         (*@\verb+`+@*)-DeclRefExpr 0x7fac9d87b828 <col:16> (*@\verb+'+@*)float(*@\verb+'+@*) lvalue ParmVar 0x7fac9d87b638 (*@\verb+'+@*)b(*@\verb+'+@*) (*@\verb+'+@*)float(*@\verb+'+@*)

\end{fancylisting}

In Listing \ref{lst:LLVM IR Example} we show our mutation example in the \LLVMIR format. Line 2 to 9 represent `\textbf{return} a *b;'
As this is closer to the machine code, less information from the original source code is preserved, e.g. the variable names \textit{a} and \textit{b} are converted to numbers \textit{\%0} and \textit{\%1}, and low-level instructions such as the allocation (alloca) of variables are introduced.
%This is achieved by letting the frontend (clang) emit the IR.
Mutating the multiplication in the \LLVMIR code means mutating the \textit{fmul} statement to \textit{fadd, fsub}, and \textit{fdiv}.
An \LLVMIR mutation testing tool usually works in two steps. It first creates a list of mutation points (specific \LLVM instruction and its operands).
It will then create a new \LLVMIR version for each mutant by applying the mutant to the mutation point.

\begin{fancylisting}{LLVM IR Example,label=lst:LLVM IR Example}
define float @_Z1fff(float %0, float %1) #0 {
	%3 = alloca float, align 4
	%4 = alloca float, align 4
	store float %0, float* %3, align 4
	store float %1, float* %4, align 4
	%5 = load float, float* %3, align 4
	%6 = load float, float* %4, align 4
	%7 = fmul float %5, %6
	ret float %7
}
\end{fancylisting}

As there are many more instructions at the \LLVMIR-level, it presents more opportunities for mutations.
However, mutations can exist in the \LLVMIR code which cannot be achieved by changing the original source code~\cite{denisov2018mull}. As no conversion to the original code is possible, there is no traceability and no easy feedback for the developers.
Additionally, not all mutation opportunities from the source code are available at the \LLVMIR-level due to optimisation in the conversion to \LLVMIR.
In short, \LLVMIR mutations provide a different set of mutants than mutating at the AST-level.

\subsection{Existing \LLVM and \Clang Mutation Testing Tools}

Below we discuss the most prominent mutation testing tools that are based on \Clang and/or the \LLVMIR.
Table \ref{table:tools} lists them in alphabetical order with their features and optimisations.
For each of them, we briefly explain which optimisations are incorporated and refer to quantitive evidence if present.
\begin{table*}[hbt]
\centering
\caption{Clang and \LLVMIR Mutation Testing Tools}
\label{table:tools}
\begin{tabular}{l|l|l|p{6,5cm}}
	\textbf{Tool Name} & \textbf{Mutation level} & \textbf{Mutation Operators} & \textbf{Mutation Optimisation} \\\hline
	AccMut & \LLVMIR & AOR, ROR, LCR, SDL, ... & Mutant Schemata, modulo states\\\hline
	CCmutator & \LLVMIR & Concurrency Mutation Operator & \\ \hline
	\multirow{2}{*}{\Dextool} & \multirow{2}{*}{AST} & \multirow{2}{*}{AOR, ROR, LCR, SDL, UOI} & Distribution, Mutant Schemata \\
	& & & Exclude unreached mutants via code coverage\\ \hline
	Mart & \LLVMIR & Operator groups & Trivial Compiler Equivalence\\ \hline
	MuCPP & AST & Class Level Mutants & Reduced Mutants Set\\ \hline
	Mull & \LLVMIR & \LLVM fragments & Limit total mutants based on call-depth\\ \hline
	\multirow{2}{*}{SRCIROR} & AST & \multirow{2}{*}{AOR, LCR, ROR, ICR} & Trivial Compiler Equivalence\\
	& \LLVMIR & & Exclude unreached mutants via code coverage\\ %\hline
\end{tabular}
\end{table*}

\myparagraph{AccMut (IR-based):} AccMut~\cite{wang2017faster} is an \LLVMIR mutation testing tool that reduces redundant execution statements anywhere in the program by analysing the original and mutated program. It identifies the redundant statements by inspecting the (local) state of both programs. When they are identical, all following statement executions are identical and redundant until the next different statement. They have demonstrated an average speedup of 8.95x over a mutant schemata approach~\cite{wang2017faster} .

\myparagraph{CCmutator (IR-based):} CCmutator~\cite{kusano2013ccmutator} is an \LLVMIR mutation testing tool specifically designed to mutate concurrency constructs.

\myparagraph{\Dextool (AST-based):} Dextool is an open-source framework created for testing and static analysis of (often safety-critical) code.
The Dextool framework is used within industry, for example within Saab Aeronautics.
 One of the plugins in the framework is \Dextool.
It was developed with a heavy emphasis on the reporting part of mutation testing in order to better understand the output of mutation testing and to gain more insight into the project under test.

\Dextool works by (textually) inserting mutants into the source code one at a time after analysing the abstract syntax tree (conveniently available via \Clang) for points to mutate.

\Dextool allows users to provide scripts and special flags in order to compile and test projects with the explicit intention to scale for more and bigger (industrial) projects.
\Dextool supports a distributed setup as the mutants are stored in a central database.
Multiple nodes can then access the database and execute a subset of all mutants.
During this study, we created a proof-of-concept schemata plugin for \Dextool to enable mutant schemata.
The steps in the \Dextool schemata plugin are identical to our standalone tool as described in this paper.
We worked with the creators of Dextool to run our experiments on the Saab Case.
The results and timings from the Saab Case were gathered using the \Dextool schemata plugin.
As a result, the creators of Dextool created a proper implementation for mutant schemata into the tool.\footnote{\url{https://github.com/joakim-brannstrom/dextool/tree/master/plugin/mutate/contributors.md}}\footnote{\url{https://github.com/joakim-brannstrom/dextool/blob/master/plugin/mutate/doc/design/notes/schemata.md}}

\myparagraph{Mart (IR-based):} Mart~\cite{chekam2019mart} is an \LLVMIR mutation testing tool that currently supports 18 different operator groups (with 68 \textit{fragments} and 816 operators).
These operator groups match against the \LLVMIR syntax to create the mutants.
Additional operator groups can be implemented by the user to further extend its capabilities.
Mart has an in-memory implementation of Trivial Compiler Equivalence to eliminate equivalent and duplicate mutants~\cite{papadakis2015trivial}.

\myparagraph{MuCPP (AST-based):}
MuCPP is a {\Clang}-based mutation testing program that generates mutants by traversing the \Clang AST and storing the mutants in different branches using a version control system~\cite{delgado2017assessment}.
MuCPP implements mutations at the class level.
These include mutations related to inheritance, polymorphism and dynamic binding, method overloading, exception handling, object and member replacement, and more.
The study is aimed at reducing the total amount of mutants that need to be executed by eliminating so-called unproductive mutants.
These include equivalent mutants, invalid mutants, easy-to-kill mutants, and mutants in dead code.
MuCPP demonstrates that \Clang can be used for implementing mutant analysis and generation at the AST level.
The study lists the speedup gained from the reduced mutation set but does not list the overhead impact of the generation and analysis of the mutants using the \Clang framework.
It does not implement other optimisation techniques, so there are no detailed measurements of the potential reduction in compilation and execution overhead using the \Clang framework, nor the overhead the implementation of such techniques using the \Clang framework might cause.

\myparagraph{Mull (IR-based):}
Mull is an open-source mutation testing tool\footnote{\url{https://github.com/mull-project/mull}} which modifies fragments of the \LLVM intermediate representation (\LLVMIR).
It only needs to recompile the modified fragments in order to execute the mutants, keeping the compilation overhead low~\cite{denisov2018mull}.
As Mull modifies \LLVM code, it is compatible with all programming languages that support compilation to \LLVMIR, such as C, C++, Rust, and Swift.
Mull includes a \textit{do-fewer} optimisation where you can limit which mutants are executed to only those mutants that are within a certain call-depth starting from the test case.
The study reports on the total runtime for the optimised mutation tool but does not provide details for the runtime of the individual steps nor is the tool contrasted to a traditional, unoptimised approach.
While the tool certainly provides a speedup for mutation testing, the lack of detailed measurements makes it difficult to estimate where the advantages lie and where overhead might occur.

\myparagraph{SRCIROR (AST or IR-based):}
SRCIROR~\cite{hariri2018srciror} is a toolset for performing mutation testing at the AST level or at the \LLVMIR level.
Both variants implement the \textit{AOR, LCR, ROR, ICR} mutation operators.
When mutating the AST, SRCIROR uses the AST matchers from the \Clang LibTooling library to search for candidate mutation locations in the AST.
It then generates a different source file for each of the generated mutants.
When mutating the \LLVMIR, SRCIROR creates a list of mutation opportunities (specific \LLVM instruction and one of its operands) within the generated \LLVMIR code of the project.
SRCIROR then creates a mutated version for each of these mutation opportunities.
SRCIROR allows to filter out unreachable mutants based on code coverage metrics.
It also allows to filter out some equivalent mutants using trivial compiler equivalence~\cite{papadakis2015trivial}.

\hypobox{\textbf{Summary.}
The current state-of-the-art demonstrates that mutant analysis for the C language family is possible on top of the \LLVM and \Clang compiler framework.
However, the extent to which the various optimisation strategies allow reduced compilation and execution overheads is unknown.
In particular, there exist no detailed measurements on two of the most advanced techniques (\textit{mutant schemata} and \textit{split-stream mutation testing}) for the C language family.}

%%%%%%%%%%%%%%%%%%%%%%%%%%%%%%%%%%%

\section{Proof-of-Concept Tool}
\label{sec::proofOfConcept}
In this paper, we investigate to which extent the \Clang front-end and its state-of-the-art program analysis facilities allow to implement existing strategies for mutation optimisation within the C language family.
For this, we implement five strategies: exclude invalid mutants, exclude unreachable mutants, mutant schemata, \hybrid, and split-stream.

The goal of these optimisation strategies is twofold.
On the one hand they try to eliminate the compilation overhead. 
On the other hand, the optimisation strategies try to reduce the execution overhead.
The different strategies build on each other, and successively introduce more optimisations.

The first compilation overhead comes from the invalid mutants, which cause compilation errors and can therefore not be executed.
Excluding them reduces the compilation overhead.
The second compilation overhead comes from individually compiling each mutant, which is computationally expensive. A mutant schemata strategy eliminates this overhead by compiling all mutants at once, drastically reducing the compilation time.

The first execution overhead comes from the unreachable mutants.
These mutants are unreachable by the test suite, hence, they will always survive.
If we know which mutants are unreachable, we can label them as survived without needing to execute them.
The second execution overhead comes form the fact that not all test cases reach each mutant.
In order for a test case to potentially kill a mutant, that test case needs to first reach and execute the target mutant.
If the test case does not reach the mutant, we know that it will never be able to kill it, hence we know its result before we execute it.
We created a detection algorithm that avoids this redundant execution by detecting which test case reaches which mutant, and thus only executes a subset of test cases for each mutant.
The final optimisation strategy we looked into is the split-stream mutation.
This strategy exploits the state-space information so that the execution of each mutant can start from the mutation point itself instead of always starting the execution at the beginning of the test suite.
This essentially cuts the amount of statements that need to be executed in half.

\begin{figure*}[b]
	\centering
	\caption{Implementation strategies with algorithm steps (first step at the bottom). Each strategy improves on the previous one by reducing compilation and/or execution overhead.}
	\label{fig:Algorithm Steps}
	\includegraphics[max size={\textwidth}{\textheight}]{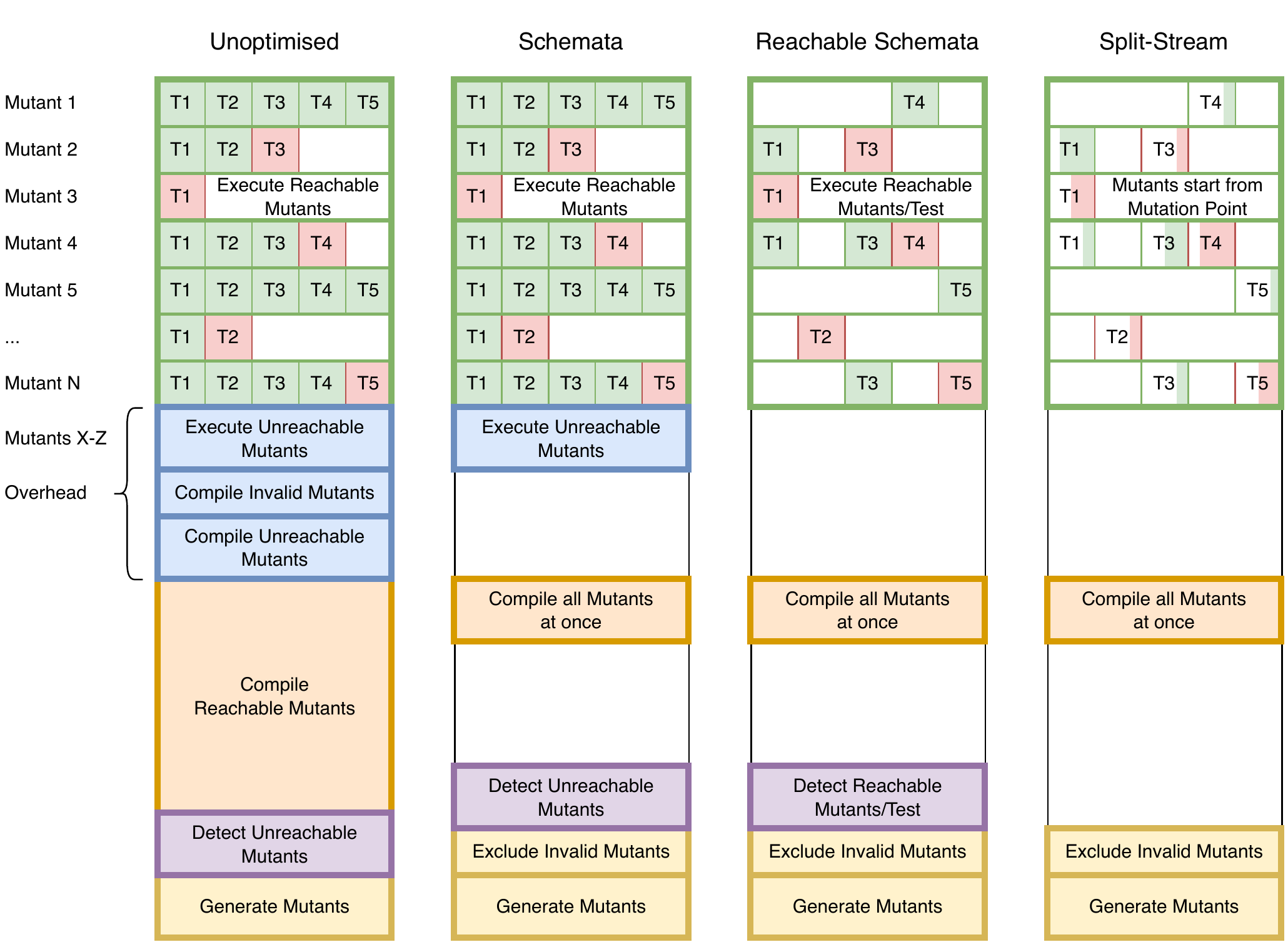}
\end{figure*}

We present a proof-of-concept optimisation tool, featuring an unoptimised approach for a baseline and the aforementioned optimisation strategies.
We discuss each strategy, their differences and similarities, the potential impact on compilation and execution time and provided detailed measurements of the steps in the optimisation strategies.
The general steps of the optimisation strategies are visualised in~\figref{fig:Algorithm Steps}. We will explain each of the optimisations by its steps, starting at the bottom of the image to the top. 

\myparagraph{Generate Mutants.}
(Bottom boxes in ~\figref{fig:Algorithm Steps}.)
The generation of the mutants is done independently of the optimisation strategies.
This allows us to use the same mutants, ensuring a fair comparison of the strategies.
This is represented by its inclusion in every pillar in~\figref{fig:Algorithm Steps}.

Currently, for our proof-of-concept tool, we have implemented the Relational Operator Replacement (ROR), Arithmetic Operator Replacement (AOR) and Logical Connector Replacement (LCR) (see \tabref{tab:mutationoperators}).
This is a representative selection of all possible mutation operators, sufficient to demonstrate the feasibility of \Clang based optimisation strategies.
We discuss this limitation under \secref{sec::Threats}.

Our proof-of-concept tool relies on the LibTooling library of \Clang, and its ability to iterate through all declarations, statements, and expressions from the abstract syntax tree (AST).
For each of these declarations, statements, and expressions, the implemented mutation operators will create corresponding mutants.
As an example, the AOR mutation operator when encountering the multiplication `$*$' operator residing in the binary expression `$a * b$' will create the mutants where the multiplication is replaced by `$+$', `$-$', `$/$', and `$\%$' operators.
To ensure that we know where to replace the original operator with the mutated one, we store the filename in which the mutant occurs, the offset to the beginning and the end of the original operator, and the mutant itself.
For each executed mutant, we store whether or not the mutant was reached, how long it took to execute, and whether or not the mutant was killed.
As it is possible that some mutants will cause an infinite loop, the user can set a maximum execution time for each mutant.
If the test suite is not able to finish executing within this time, we stop the test execution and say that the mutant \textit{timed out}.
Mutants that are timed out can be considered as killed, as they would also time out in a continuous integration test.
This information can then be extracted to form a report to inform developers where and which mutants survived the test suite.

\subsection{Unoptimised Mutation Testing}
In a traditional, unoptimised mutation testing setting each mutant is inserted, compiled and executed separately.
Some mutants, however, can be located in uncovered code.
Such mutants are unreachable and are unable to infect the program state, thus they can never be killed.
In a realistic scenario, one would first run a code coverage technique to determine which mutants are located in uncovered code and instantly label them as survived.
As the purpose of this paper is to gain more insight into the potential speedups of our optimisation techniques, we do generate and execute them in order to gain more insight into the kind of overheads these unreachable mutants introduce.

For detailed measurements, we timed the generation of the mutants, the compilation of the mutants, and the execution of the mutants.
For the mutants themselves, we make a distinction between the invalid mutants (i.e. mutants that cause compilation errors) and the valid mutants. We further divide these valid mutants into the unreachable mutants (i.e. mutants that no test case reaches), and the reachable mutants.

We implemented the unoptimised mutation testing approach to obtain a baseline for the mutation analysis and to verify the correctness of our optimisation strategies.

\myparagraph{Detect Unreachable Mutants.}
In order to determine which mutants are completely unreachable by the test suite, we instrument the code base with a wrapper on the mutated locations.
Running the test suite with the instrumented code base then provides us with a list of mutants that are reachable by the test suite.
From this, we can deduce the completely unreachable mutants.

\myparagraph{Compile Mutants.}
After the generate mutants phase, each mutant needs to be individually inserted into the original code base and compiled before it can be executed.
We measured the compilation time for each mutant and divided them into three categories: reachable, unreachable, and invalid mutants.
This allows us to quantify the overhead caused by the unreachable and invalid mutants.

All three can be seen in the first pillar, i.e. unoptimised, in~\figref{fig:Algorithm Steps}.
It is important to note that we do not clean the build environment between the different mutants.
We use consecutive builds to speed up the compilation, as the compiler then only needs to re-compile the files that are impacted by the mutant.

\myparagraph{Execute Mutants.}
After generation and compilation, the valid mutants need to be executed.
For each mutant, the entire test suite needs to be run.
The test cases of the test suite are represented by T1 to T5 in \figref{fig:Algorithm Steps}.
When a test case fails, represented by the red colour, this means that the mutant is killed.
When none of the test cases fail, the mutant survives, represented by the green colour.
To optimise the mutation testing execution, we rely on the early-stop principle, for which the test suite execution stops when the first test case kills the mutant.
In the unoptimised approach, the entire test suite is run for each unreachable mutant. We represented this with mutants X-Z in \figref{fig:Algorithm Steps}.

\subsubsection*{Expected Performance}
For the unoptimised approach, we estimate the performance via Formula \ref{eq:unoptimised}.
The formula consists of three phases: the generate mutants, compile mutants, and execute mutants phase.
We included the generation of the mutants in the formula as this step is necessary and identical for all approaches but its impact should be negligible.
We do not include the detection of unreachable mutants as it is not part of an unoptimised approach. We only need it to distinguish between the unreachable and reachable mutants.
The total compilation time is subject to the amount of reachable, unreachable and invalid mutants.
From our measurements, we have seen that invalid mutants can take up to 10\% of the total execution time.
The total test suite execution time is only subject to the amount of reachable and unreachable mutants as invalid mutants cannot be executed.
Note that the compilation time of consecutive builds is lower than a clean build and that the test suite execution time varies depending on where, if at all, the mutant is killed due to the early-stop mechanism.

\begin{equation}
\label{eq:unoptimised}
\begin{tabular}{rll}
				&$t_{\textit{mutant\_generation}}$	&\\
+\hspace{-1em}	&$t_{\textit{compilation}}$			& $*(\textit{reachable\_mutants} + \textit{unreachable\_mutants} + \textit{invalid\_mutants})$\\
+\hspace{-1em}	&$t_{\textit{test\_suite\_execution}}$			& $* (\textit{reachable\_mutants} + \textit{unreachable\_mutants})$
\end{tabular}
\end{equation}

\subsection{Mutant Schemata}
The mutant schemata strategy compiles all mutants at once, essentially eliminating the compilation overhead~\cite{untch1993mutation}.
Previous studies with mutant schemata have shown that this optimisation approach can provide an order of magnitude improvement on a full mutation analysis.
Untch et al.
%\citeauthor{untch1993mutation} 
reported a preliminary speedup of 4.1~\cite{untch1993mutation}.
In a later study, Wang et al. confirmed these findings and reported a speedup between 6.46 and 14.00 on systems written in Java~\cite{wang2017faster}.
However, we found no specific studies investigating the speedups of mutant schemata for the C language family, hence, in this study, we will collect detailed measurements.
We timed the generation of the mutants, the compilation of all the mutants at once, and the execution of the mutants.
We can then compare this to the unoptimised approach to gain insights into its speedups and potential overheads.
For a fair and complete comparison, we again make the distinction between the unreachable and reachable mutants.

\myparagraph{Exclude Invalid Mutants.}
As all mutants are inserted into a single compilation unit, all generated mutants need to compile.
If even a single mutant causes a compilation error, the complete mutation analysis will fail.
This is especially challenging for statically typed languages with many interacting features and unforgiving compilers (C, C++, \ldots).
This is where the \Clang compiler front-end can be used for ensuring that all injected mutants will compile.
As \Clang has access to all the type information from the project, our program can verify, during the generate mutants phase, that a newly created mutant is compile-time correct.
For this, we rely on the semantic analyser of \Clang.
As all the information is available while traversing the AST, the additional analysis time should be limited.
If a mutant is incorrect, for example ``string - string'' or ``float \% int'', the mutant is labelled as invalid, as the mutant would cause a compilation error.
To ensure that we exclude only the invalid mutants, we verified that the mutants we labelled as invalid are the same mutants that actually caused compilation failures in the unoptimised baseline.
As this step is tightly integrated with the generation of the mutants, we measured this together with the generation of the mutants.

\myparagraph{Detect Unreachable Mutants.}
The detection of the unreachable mutants is identical to the one for the unoptimised approach.

\myparagraph{Compile Mutants.}
Instead of compiling each mutant separately, the mutant schemata strategy compiles all mutants at once.
This means that all mutants are inserted into the code, each mutant being guarded by a conditional statement allowing the activation of individual mutants at run-time.
The result is a single code base and only one compilation is needed, drastically reducing the compilation time.
We visualised this in~\figref{fig:Algorithm Steps} by using a smaller `compile all mutants at once' block .
A specific mutant is activated by using an external environment variable.
For this, additional code, i.e. the external variable, needs to be added to each file of the project (see line 1 in Listing \ref{lst:Mutant Schemata Example}).
Listing \ref{lst:Mutant Schemata Example} shows the mutant schemata version of the original ``$a * b$'' example.
We use the ternary operator, the short-handed version of an \emph{if} statement, to allow nesting the mutants inside the condition of \emph{if} statements.

In the example, the `$\%$' operator gives an ``\texttt{invalid operands to binary expression}'' error, therefore, we cannot compile our program unless we remove this mutant.
As we exclude invalid mutants during generation, these will not show up in the mutated code base.

\begin{fancylisting}{Mutant Schemata Example,label=lst:Mutant Schemata Example}
extern int MNR;	// prepended external variable, allowing selection of active mutant

// mutated ``return a * b;'' statement using the ternary operator:
float f(float a, float b) { 
	return (MNR == 1 ? a + b :
		   (MNR == 2 ? a - b :
		   (MNR == 3 ? a / b :
		   (MNR == 4 ? a % b : a * b))));
                      (*@ \textasciitilde\textasciitilde\textasciitilde \textasciicircum\textasciitilde\textasciitilde\textasciitilde \textcolor{red}{Invalid operands to binary expression} @*)
}
\end{fancylisting}

\myparagraph{Execute Mutants.}
As a final step, all the mutants need to be executed.
This can be done by running the executable for each valid mutant and only changing the environment variable.
We again make the distinction between the reachable mutants, represented by mutants 1 to N, and the unreachable mutants represented by mutants X-Z in \figref{fig:Algorithm Steps}.

\subsubsection*{Expected Performance}
For the mutant schemata approach, we estimate the performance via Formula \ref{eq:schemata}.
The formula consists of three phases: the generate mutants, the compile mutants, and execute mutants phase.
We included the generation of the mutants in the formula as this step is necessary and identical for all approaches but its impact should be negligible.
We do not include the detection of unreachable mutants as it is not part of the schemata approach.
We only need it to distinguish between the unreachable and reachable mutants.
As the strategy removes the compilation overhead by only compiling the project once, instead of for each mutant, a drastic speedup is to be expected.
Therefore, we no longer need to multiply the compilation time by the amount of mutants.
Note that the compilation time will be slightly longer, as there is more code that needs to be compiled.
The execution part of the formula stays the same, as we still need to execute the test suite for each mutant. 
The test suite execution time for each mutant will, however, also be slightly longer as there is more code that needs to be executed due to the \emph{if} statements of the ternary operator in order to activate the correct mutant.

\begin{equation}
\label{eq:schemata}
\begin{tabular}{rll}
				&$t_{\textit{mutant\_generation}}$				&\vspace{0.5em}\\
+\hspace{-1em}	&$t^{\textit{schemata}}_{\textit{compilation}}$			&\vspace{0.5em}\\
+\hspace{-1em}	&$t^{\textit{schemata}}_{\textit{test\_suite\_execution}}$	& $*(\textit{unreachable\_mutants} + \textit{reachable\_mutants})$
\end{tabular}
\end{equation}

\subsubsection*{Schemata Implementation}
\label{sec:Schemata Implementation}
We implemented the mutant schemata technique using the ternary operator, the short-handed version of an \emph{if} statement.
We do note that the most common case for this implementation is the worst-case scenario.
Most of the times no mutant in the statement will be activated, causing an evaluation of all the if statements before the original statement can be executed.
We believe that this will create a fixed overhead per mutant, but we expect this overhead to be limited.
The advantages of this approach is that it simplifies and streamlines the implementation of the schemata technique.
It allows straightforward nesting the mutants inside the condition of an \emph{if} statement.
This can be seen in Listing \ref{lst:Mutating if Statements}.
The body of the if statement can also directly be mutated inside the expression. 

\begin{fancylisting}{Mutating \textit{if} Statements using ternary operator,label=lst:Mutating if Statements}
// original
if (a > b) {/*body*/}

// mutated using ternary operator
if	(MNR == 1 ? a <  b :
 	(MNR == 2 ? a <= b :
 	(MNR == 3 ? a == b :
 	(MNR == 4 ? a >= b : a > b)))) {/*mutated body*/}
\end{fancylisting}

\myparagraph{\textbf{An alternative implementation}} is possible by using a switch case approach.
We can nest the switch inside the condition of the if statement using statement expressions\footnote{\url{https://gcc.gnu.org/onlinedocs/gcc/Statement-Exprs.html}}, as seen in Listing \ref{lst:Mutating switch}.
While statement expressions are supported by default in gcc and clang, they are not supported by the C standard.
We are thus strongly discouraged to use them if we want to enable mutation testing in an industrial setting where it is not always guaranteed that the used compiler will support the GNU extension.
The easiest workaround is putting the statements in a function and calling said function in the condition of the if statement. For each mutated statement, a new function will then have to be created.

We can also implement the schemata technique using a global switch, which has its own set of challenges.
Firstly, switch cases have a local scope.
This means that when mutating an assign statement using the switch case approach, the variable will first need to be defined outside the scope of the switch case.
Secondly, a code explosion will occur as the body of the if statement needs to be repeated for each case, see Listing \ref{lst:Mutating switch}.
The code explosion can be reduced by creating functions for the body.

We can then either mutate the body inside the default case using additional switch cases, or append the cases of the mutated body to the existing switch.
The latter reduces the amount of switch cases, but would cause a code explosion.

\begin{fancylisting}{Mutating \textit{if} Statements using switch-cases,label=lst:Mutating switch}
// original
if (a > b) {/*body*/}

// mutating using nested switch case
if (({bool r;
	  switch (MNR) {
          case 1:   r = a < b;  break;
          case 2:   r = a <= b; break;
          case 3:   r = a == b; break;
          case 4:   r = a >= b; break;
          default:  r = a > b;
      }
      r;}){/*mutated body*/}
}

// mutating using global switch case
switch(MNR) {
	case 1:  if (a >  b) {/*body*/} break;
	case 2:  if (a >= b) {/*body*/} break;
	case 3:  if (a == b) {/*body*/} break;
	case 4:  if (a >= b) {/*body*/} break;
	default: if (a >  b) {/*mutated body*/} break;
}
\end{fancylisting}

\subsection{Reachable Schemata}

The previous strategy essentially eliminated the compilation overhead.
This causes the test suite execution to become the most time-consuming part of the mutation analysis.
While eliminating the completely unreachable mutants does speed up the mutation testing analysis, the \textsf{reachable schemata strategy} aims to use a more fine-grained approach to reduce most of the execution overhead.
The \textsf{reachable schemata strategy} reduces the test suite scope to only those test cases which reach the mutant, effectively reducing the execution overhead.
We timed the generation of the mutants, the compilation of all the mutants at once, and the execution of a reduced test suite for each of the mutants.
We can then compare this to the unoptimised and mutant schemata strategies to gain insights into its speedups and potential overheads.

\myparagraph{Exclude Invalid Mutants}
As the reachable schemata strategy is an optimisation based on the regular mutant schemata approach, we need to exclude the invalid mutants as well.
This is done in the same way as with the mutant schemata approach.

\myparagraph{Detect Reachable Mutants/Test}
In order to determine which mutants are reached by which test cases, we instrument the code base with a wrapper on the mutated locations.
We store each mutated location a test run reaches.
Running the instrumented code base for each test case provides us with lists of mutants that are reachable by each test case.
This results in a reduced set of test cases which need to be executed for each mutant, resulting in a speedup without information loss.

The best results are obtained by using fine-grained test selection and running each test case individually.
The speedups of this technique will depend on the test driver used for each project.
Some test drivers might only be able to run individual modules instead of individual test cases.
Speedups from module-grained test selection will be lower than from fine-grained test selection.
Our projects were all compatible with fine-grained test selection on a test-by-test case basis.

\myparagraph{Compile Mutants}
The compilation of the mutants is identical to the one for mutant schemata.
We do note that an extra optimisation can be done by only inserting and compiling the reachable mutants in this phase. We, however, did not yet implement this optimisation in our proof-of-concept.

\myparagraph{Execute Mutants}
As a final step, all the mutants need to be executed.
Instead of running the entire test suite for each mutant, we now only need to run those test cases which reach the mutated locations. 
In ~\figref{fig:Algorithm Steps} we see that mutant 1 is not reachable by the entire test suite, therefore no tests are executed for the mutant.
For mutant 2, the second test case is not executed as that test case does not cover mutant 2.

\subsubsection*{Expected Performance}
For the reachable schemata approach, we estimate the performance via Formula \ref{eq:reachable schemata}.
The formula consists of four phases: the generated mutants, the detect reachable mutants, the compile mutants, and the execute mutants phase.
We included the generation of the mutants in the formula as this step is necessary and identical for all approaches but its impact should be negligible.
We included the detection of the reachable mutants/test as it is a part of the \hybrid approach in order to gather the reachable mutants to the test case relationship.
We estimate its time impact at a single compilation and execution of the test suite.
This is immediately compensated when there is more than one unreachable mutant.
The compilation part again consists of a single compilation as it is an extension of the mutant schemata approach.
For the execution part, only the reachable mutants are considered, together with a decoupling factor. This factor is a percentage based on the reduction in average mutants reachable by each test case.
The more coupled a project is, the less effective this strategy will be.
The less coupling there is (especially in combination with mocking), the more effective this strategy will be.
From our measurements, we have seen that the average amount of mutants reached per test is between 10 and 20\% of all valid mutants.
This implies a speedup between 5 and 10x.

\begin{equation}
\label{eq:reachable schemata}
\begin{tabular}{rl}
				&$t_{\textit{mutant\_generation}}$				\vspace{0.5em}\\
+\hspace{-1em}	&$\left(t^{\textit{schemata}}_{\textit{compilation}} + t^{\textit{schemata}}_{\textit{test\_suite\_execution}}\right)$			\vspace{0.5em}\\
+\hspace{-1em}	&$t^{\textit{schemata}}_{\textit{compilation}}$			\vspace{0.5em}\\
+\hspace{-1em}	&$t^{\textit{schemata}}_{\textit{test\_suite\_execution}}$	 $*\textit{reachable\_mutants} * {\textit{decoupling\_factor}}$
\end{tabular}
\end{equation}

\subsection{Split-Stream Mutation Testing}
The split-stream mutation testing strategy reduces the execution overhead of mutation testing even further.
Instead of letting each mutant initiate execution from the start of the program, each mutant is started from the mutation point itself.
This can be achieved by exploiting the state-space information. Previous research has shown an average speedup of 3.49x of split-stream mutation testing over mutant schemata by mutating the \LLVMIR~\cite{wang2017faster}.

\begin{figure}[hbt]
	\centering
	\caption{Trace of Original and Mutated programs}
	\label{fig:Trace of Original and Mutated programs}
	\includegraphics[width=0.8\linewidth]{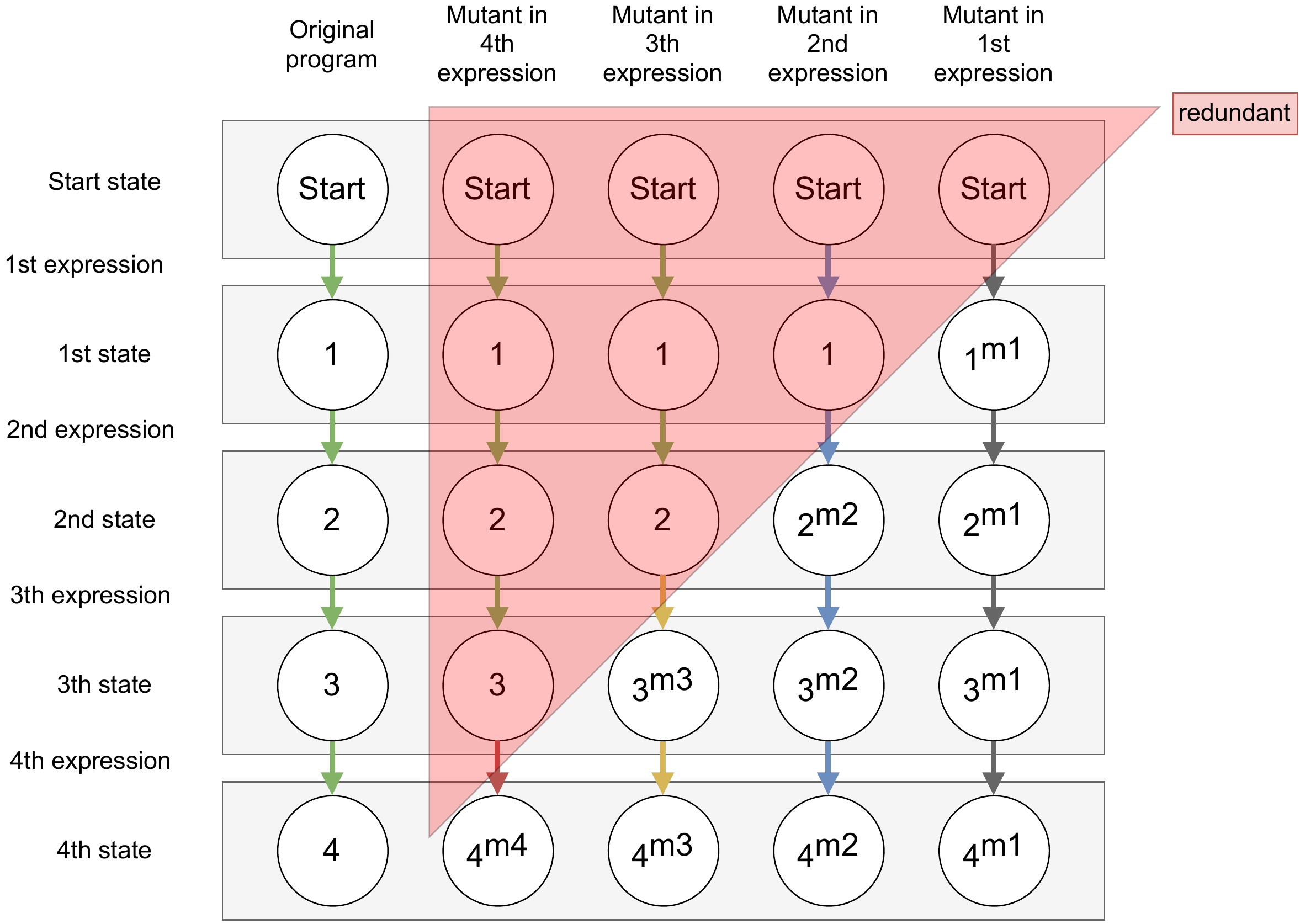}
\end{figure}

We explain the overhead and general idea of the split-stream mutation testing approach using \figref{fig:Trace of Original and Mutated programs}.
Here we visualised the execution trace of the original, unchanged program, as a long vertical trace of states, each representing the execution of a single instruction.
The trace of each mutant will look identical to the original mutant up until the mutated expression.
In \figref{fig:Trace of Original and Mutated programs} we show mutant 4 with a mutation in the 4th expression, mutant 3 with a mutation in the 3rd expression, ...
The trace will only start to deviate from the original trace after the mutated expression.
All of the states up until the mutated state are in fact redundant, as we already know them from the execution of the original program.
We highlighted these redundant states and expressions in red in \figref{fig:Trace of Original and Mutated programs}.

\myparagraph{Exclude Invalid Mutants}
As this strategy is an optimisation from the regular mutant schemata approach, we need to exclude the invalid mutants as well.
This is done identically as with the mutant schemata approach.

\myparagraph{Compile Mutants}
The compilation of the mutants is identical to the one for mutant schemata.
However, additional code needs to be instrumented in the code base in order to exploit the state space and start the mutants from their mutation point instead of from the start of the program.

\begin{figure*}[hbt]
	\centering
	\caption{Split-Stream Mutation process}
	\label{fig:Split-stream Mutation process}
	\includegraphics[width=0.8\linewidth]{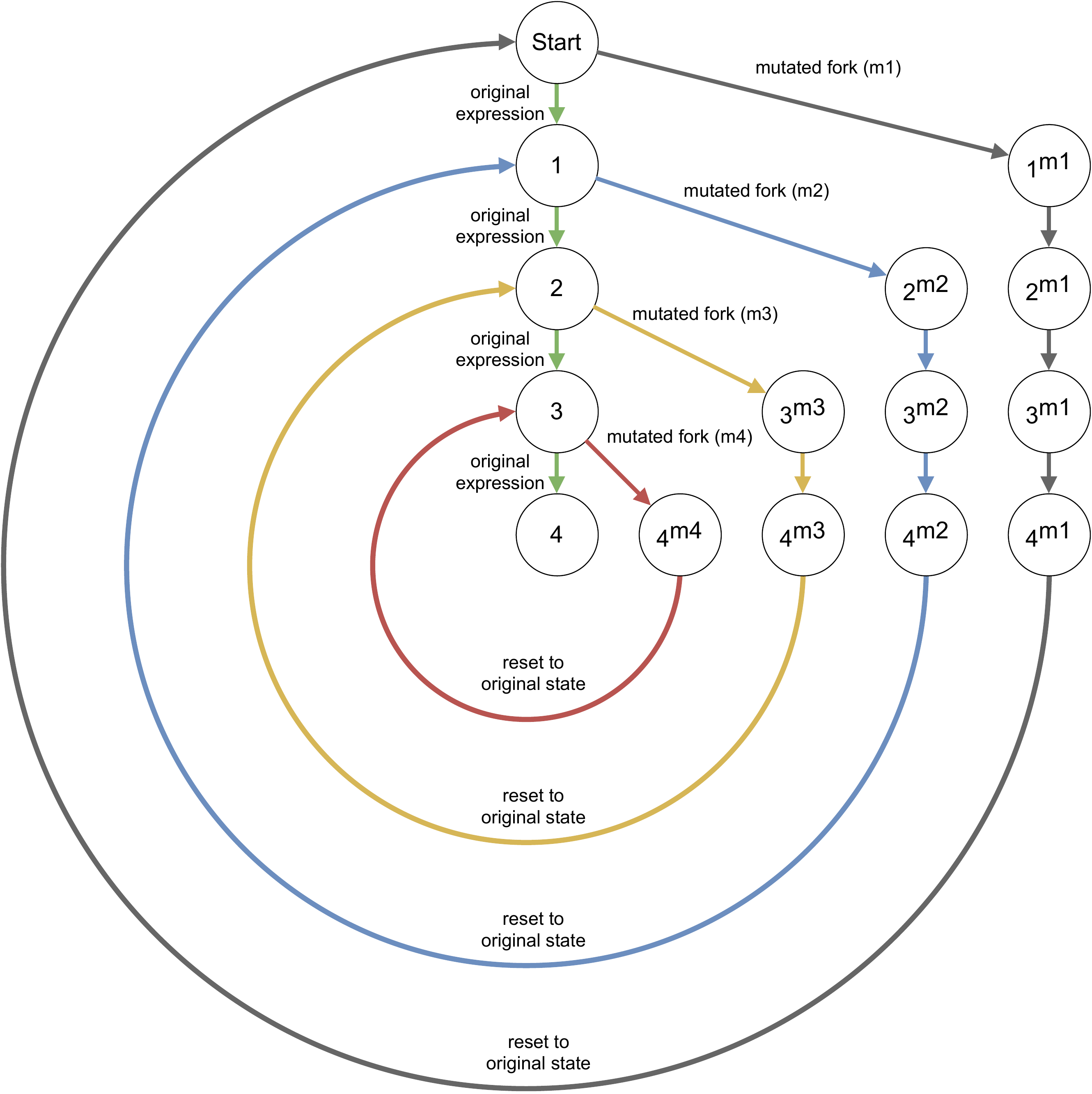}
\end{figure*}

\myparagraph{Execute Mutants}
In order to exploit the state space, we start the split-stream mutation analysis only once instead of for each mutant as with mutant schemata.
We will explain the split-stream mutation testing process using \figref{fig:Split-stream Mutation process}.

When we start the split-stream mutation analysis, no mutant is active.
The first mutant the program will encounter will not yet have been activated, so the program will fork the entire program and pause the original one.
The forked program will be in the exact same state as the original program, essentially this is a duplicate of the state-space.
In the forked program, we activate the encountered mutant.
We let the forked program execute, only taking into account the active mutant and store its results at the end of its execution.
We then continue our first program, until it encounters another mutant that it has not yet activated.
Here the process is repeated, a fork is created, it is executed, and its results are stored.
This goes on until the original program reaches its end state.

Like with the mutant schemata strategy, our proof-of-concept tool needs to decide what needs to happen at each mutation position.
Instead of only deciding whether or not a mutant needs to be active or not, it now also needs to decide \textit{when} it needs to fork the process.
For this, we instrumented additional code at each mutation point.
We explain the general concept using the C++ example in Listing \ref{lst:Split Stream Example}.

When we start our program with all the mutants in it, no mutant will ever be activated in it.
The original program is only responsible for forking newly encountered mutants and storing their results.

When a mutant is encountered in the main program, e.g. on line 16, then it will continue to line 7, where it will verify whether or not it already encountered that mutant.
If it did, it will continue the program, otherwise, it will fork the program and wait for the forked program to finish, on lines 8 and 9.
The forked program will only take into account the active mutant and run till its end.
The main program will then store the result (exit code) of the forked program, add the mutant to the already executed list and continue until it encounters a new mutant, on lines 9 and 10.

\begin{fancylisting}{Split-Stream Mutation Testing Example,label=lst:Split Stream Example}
extern list<int> MNR_list;	// already activated mutants
extern int MNR;				// active mutant

bool split(int mnr) {
	if (/* this is the fork */) { return mnr == MNR; }
	else { /* this is the main program */ 
		if (MNR_list.contains(mnr)) {return false;}
		else { /* fork, and set MNR to mnr in forked process*/ }
		/* wait for fork to complete and store results */
		MNR_list.insert(mnr);
	}
	return false;
}

float f(float a, float b) {
	return (split(1) ? a + b :
		   (split(2) ? a - b :
		   (split(3) ? a / b : a * b)));
}
\end{fancylisting}

The split-stream mutation testing approach naturally ensures that unreachable mutants are not executed.
The approach will never create forks for mutants that it will not reach.
By executing each of the tests separately, only those mutants reachable by the test case will be executed for that test.
This is similar to the reachable schemata approach, with the added benefit that there is no separate compilation and/or test suite execution run necessary in order to extract the reachable mutants.

As we start the mutants from their mutation points, we now also need to ensure that all of the project dependencies, e.g. files and databases, are in their correct state.

As this will likely require specific project knowledge and will differ for each project, we built in hook functions which can be specialised for each project.
Additionally, we built in support to automatically reset local files to the state they were in for each of the mutation points.

\subsubsection*{Expected Performance}
For the split-stream mutation approach, we estimate the performance via Formula \ref{eq:split-stream}.
The formula consists of three phases: the generate mutants, the compile mutants, and the execute mutants phase.
We included the generation of the mutants in the formula as this step is necessary and identical for all approaches but its impact should be negligible.
We no longer need to detect the reachable mutants/test as it is inherent to the split-stream approach.
The compilation part again consists of a single compilation, which should be slightly longer than the compilation of the schemata approach.
The execution time will be impacted negatively as the forking of the process will take some time.
As the forking process works on a copy-on-change principle, only the memory that is changed will be copied.
This ensures that the impact of the forking is kept at a minimum.
The benefit of this strategy is that we no longer need to execute the redundant code.
A mutant that is located at the very end of the execution will now only take very little time to execute.
This is in contrast to the original mutation testing where it would have needed to start its execution from the very beginning.
Mutants located in the front will only have a small improvement in their execution time.
In general, one could thus assume that this strategy would contribute to an additional 2x speedup.
Similar to the \hybrid approach, executing the test suite on a test-by-test case will yield the best results. Hence, we have the same decoupling factor.

\begin{equation}
\label{eq:split-stream}
\begin{tabular}{rl}
				&$t_{\textit{mutant\_generation}}$				\vspace{0.5em}\\
+\hspace{-1em}	&$t^{\textit{split\_stream}}_{\textit{compilation}}$			\vspace{0.5em}\\
+\hspace{-1em}	&$t^{\textit{split\_stream}}_{\textit{test\_suite\_execution}}*\textit{reachable\_mutants} * \textit{decoupling\_factor}/2$
\end{tabular}
\end{equation}

%%%%%%%%%%%%%%%%%%%%%%%%%%%%%%%%%%%%%%%%%%

\section{Experimental Set Up}
\label{sec::Exp_setup}

\renewcommand*{\RQOne} {How much speedup can we gain from mutant schemata?}
\renewcommand*{\RQTwo} {How much speedup can we gain from split-stream mutation testing?}
\renewcommand*{\RQThree} {How much speedup can we gain from eliminating invalid mutants?}
\renewcommand*{\RQFour} {How much speedup can we gain from eliminating unreachable mutants?}

This paper presents a feasibility study, investigating to which extent the \Clang front-end and its state-of-the-art program analysis facilities allow to implement existing strategies for mutation optimisation within the C language family.
We measure the speedup from two perspectives (compilation time and execution time) assessing four optimisation strategies.
This gives rise to the following research questions:
\begin{compactitem}
	\item \textit{RQ1: \RQOne}
	\item \textit{RQ2: \RQTwo}
	\item \textit{RQ3: \RQThree}
	\item \textit{RQ4: \RQFour}
\end{compactitem}

To answer these research questions, we need to measure the impact of our optimisations, which delays we introduce, how much of the overhead we eliminate, and more importantly where these strategies can be applied.

\subsection{Cases}
To investigate the strengths and weaknesses of the \Clang-based optimisation strategies, we validate our proof-of-concept tool on four open-source C++ libraries and one industrial component.
These cases cover a wide diversity in size, C++ language features used, compilation times, and test execution times as shown in \tabref{table:ProjectDetails}.
To allow other researchers to reproduce our results, we refer to the latest commit id of the version of the project we used for our analysis.
Our tools are available in an open-source GitHub repository\footnote{\url{https://github.com/Sten-Vercammen/F-ASTMut.git}}.

\begin{table*}[hbt]
\centering
\caption{\protect{Results: Project Details}}
\label{table:ProjectDetails}
\begin{tabular}{|l|r|r|r|r|r|r|}
  \hline
	  & TinyXML2 & JSON  & Google Test & CppCheck & Saab Case  \\\hline
	Commits & 1,052 & 4,312 & 3,840 & 25,309 &  \multirow{6}{*}{Confidential} \\\cline{0-4}
	Contributors & 78 & 213 & 100 & 340 & \\\cline{0-4}
	GitHub stars & 3.9k & 28.4k & 24.9k & 3.9k &  \\\cline{0-4}
	LOPC & 3,542 & 10,955 & 32,630 & 98,171 &  \\\cline{0-4}
	LOTC & 1,885 & 26,586 & 28,064 & 159,780  &  \\\cline{0-4}
	Test Cases & 1 & 88 & 61 & 3,745 &  \\\hline \hline
	Compilation time & 1.12s & 3m 52.88s & 6m 31.34s & 5m 49.68s &  3m 08.52s \\\hline
	Test suite run time & 0.11s & 14m 10.76s & 15.01s & 17.04s & 2.63s \\\hline \hline 
	Generated Mutants & 1,038 & 3,764 & 4,488 & 61,007 &  \multirow{4}{*}{Confidential} \\\cline{0-4}
	\makecell[l]{Excluded Mutants (Const,\\Constexpr or Templated Mutants)} & 185 & 3,254 & 1,755 & 5,591 &  \\\cline{0-4}
	Considered Mutants & 853 & 510 & 2,733 & 55,416 &  \\\hline
	Mutants/LOPC & 0.241 & 0.047 & 0.084 & 0.564 &  0.233 \\\hline\hline

	Valid Mutants & 716 & 333 & 2,498 & 54,643 & \multirow{5}{*}{Confidential} \\\cline{0-4}
	\makecell[l]{Invalid Mutants (killed by compiler)} & 137 & 177 & 235 & 773 &  \\\cline{0-4}
	Completely Unreachable Mutants & 36 & 0 & 144 & 5,712 &  \\\cline{1-5}
	\multirow{2}{*}{AVG Tests/Reachable Mutant} & 0.95 & 8.46 & 11.91 & 402.96 &  \\\cline{2-5}
	& 94.97\% & 9.61\% & 19.53\% & 10.76\% &  \\\hline \hline
	Survived Mutants & 331 & 33 & 1,028 & 21,291
	&  \multirow{3}{*}{Confidential} \\\cline{0-4}
	Killed Mutants (excl.
timed out) & 211 & 300 & 1,419 & 24,639
	&   \\\cline{0-4}
	Timed out & 138 & 0 & 51 & 8,713
	&   \\\hline	
\end{tabular}
{\\\footnotesize LOPC = Lines of Production Code; LOTC = Lines of Test Code.
LOPC (incl.
include files) calculated using cloc\footnotemark (excl.
newlines and comments)}
\end{table*}
\protect\footnotetext{\url{http://cloc.sourceforge.net}}

In the first block of the table, we list general details about the project, number of commits, contributors, Lines Of Project Code (LOPC), Lines Of Test Code (LOTC), and the number of test cases the project has.

The second block list the compilation time of the project and the test suite execution time.
Projects with a longer compilation time might benefit more from a mutant schemata strategy than projects with a longer test suite execution time.

In the third block, we list how many mutants we generated for each project.
As our mutant schemata optimisation cannot yet work with mutants in so-called const-expression and/or templates, we excluded these mutants for a fair comparison.
We also listed the amount of mutants per line of production code (LOPC) which indicates how densely or sparsely packed the mutants are.
A project with a high density of mutants might introduce delays specific to the optimisation strategy.

In the fourth block, we list the amount of valid mutants and the amount of invalid mutants, i.e. mutants that are killed by the compiler.
We also list the amount of mutants that are not reachable by the current test suite.
Finally, we list the average amount of test cases that actually reach a valid mutant. We also listed thin in percentages to show how much of the total test cases are considered for each reachable mutant.
The reachable mutant schemata strategy uses this information to its advantage to reduce the mutant execution time.

In the last block, we list the amount of survived mutants and the amount of killed mutants.

\subsubsection{TinyXML2}

\begin{tabular}{p{0.35\linewidth}l}
	\url{https://github.com/leethomason/tinyxml2} &
	\footnotesize{commit id: ff61650517cc32d524689366f977716e73d4f924}
\end{tabular}

TinyXML2 is a simple, small, efficient, C++ XML parser that can be easily integrated into other programs.
It represents our small-sized project with 3,542 lines of production code and 1,885 lines of test code (without empty lines and comments).
Even though the project is small, we generate 1,038 mutants for it.
We choose this project deliberately because of its short compilation and execution time (1.12s and 0.11s respectively), as it represents a worst-case scenario for any overhead introduced by the optimisation strategy.

\subsubsection{JSON}

\begin{tabular}{p{0.35\linewidth}l}
	\url{https://github.com/nlohmann/json} &
	\footnotesize{commit id: 7c55510f76b8943941764e9fc7a3320eab0397a5}
\end{tabular}

JSON is a special case as the entire source code consists of a single header file.
This means that any changes to that file will cause a complete rebuild of the entire project, including virtually all test files.
Furthermore, all valid mutants are reachable by the test suite.
This means that there will be no speedup from detecting the unreachable mutants.

The test suite of JSON consists of 88 tests with a runtime of 14 minutes.
Two of the tests however cause the majority of this time, i.e., test 12 with 1 minute and test 80 with 11 minutes execution time.
By limiting the number of mutants to only those mutants that are reached by these tests, the total runtime can be reduced drastically.
 
\subsubsection{Google Test}

\begin{tabular}{p{0.35\linewidth}l}
	\url{https://github.com/google/googletest} &
	\footnotesize{commit id: f2fb48c3b3d79a75a88a99fba6576b25d42ec528}
\end{tabular}

Google Test represents our medium size project with 32,630 lines of production code and 28,064 lines of test code.
It is a widely used framework for testing C++ code, so one could expect it to be fairly reliable.
This is confirmed by our mutation coverage, from the 2,498 valid mutants, only 144 (5.8\%) are unreachable by the test suite.

\subsubsection{CppCheck}

\begin{tabular}{p{0.35\linewidth}l}
	\url{https://cppcheck.sourceforge.io} &
	\footnotesize{commit id: 8636dd85597acdc1560f7e0bd364c94851bec3b9}
\end{tabular}

CppCheck is a static analysis tool for C/C++ programs.
It aims to detect bugs, undefined behaviour, and dangerous coding constructs.
It has the most mutants per line of production code and will thus also have the largest negative impact on the duration of its test suite execution.
It is the biggest of the projects we analysed with the highest amount of mutants.
We would like to note that this project has many configurable macros. We ran the project without changing the defaults. Running the project with different macro configurations can lead to different results. This can even increase or decrease the number of unreachable mutants.

\subsubsection{Saab Case}

\begin{tabular}{p{0.85\linewidth}l}
	\url{https://saabgroup.com/about-company/organization/business-areas/} &
	\footnotesize{~}
\end{tabular}

The \ifthenelse{\boolean{anonymous}}{Industrial case} {Saab Case} represents the project from our industrial collaboration.
The company develops safety-critical systems and must adhere to 100\% MC/DC testing (Modified Condition/Decision Coverage, the coverage criterion adopted for the highest Design Assurance Level (DAL) in accordance to the \ifthenelse{\boolean{anonymous}}{\textsf{anonymised}}{RTCA-DO178B/C} standard).
This means that their project is well tested.
Due to the project being classified as confidential, some information has been left out and marked \textit{Confidential} in the coming sections.
However, the information related to the speedup caused by the use of mutant schemata can be disclosed.

\subsection{Hardware and Software Set-up}
We used the same infrastructure for the analysis of the selected open-source projects.
We used an Intel(R) Core(TM)2 Quad Q9650 CPU, with two 4GB (Samsung M378B5273DH0-CH9) DDR3 RAM modules (for a total of 8GB) and a 250GB Western Digital (WDC WD2500AAKX-7) hard drive.
The PC was running Ubuntu 18.04.1 LTS (GNU/Linux 4.15.0-29-generic x86\_64).
Using an SSD will drastically influence the compilation times and negatively impact the total speedups.
Using more RAM and/or a faster CPU might influence the compilation time and/or execution time, this however is presumed to be marginal.

The compilation of the projects was done using the Clang compiler with the configuration and optimisations (flags) provided by each of the projects their configuration files.

The industrial project of \ifthenelse{\boolean{anonymous}}{our industrial partner} {Saab Aeronautics} ran on their build server. Some of their results could be slightly impacted by background services.
%%%%%%%%%%%%%%%%%%%%%%%%%%%%%%%%%%%%%

\section{Results and Discussion}
\label{sec::Results_Discussion}
Before we measure the speedup induced by the different optimisation strategies implemented using the \Clang front-end, we first analyse the time spent in each of the optimisation phases.
The individual timing results can be found in \tabref{table:IndividualTimings}.
In essence, this timing information allows us to assess the significant terms in the performance estimation formulas for the
	unoptimised configuration (\formularefpage{eq:unoptimised});
	the mutant schemata (\formularefpage{eq:schemata} and \formularefpage{eq:reachable schemata})
	and the split-stream mutation testing (\formularefpage{eq:split-stream})

\begin{table*}[!hbt]
\centering
\caption{\protect{Results: Individual Timings}}
\label{table:IndividualTimings}
\begin{tabular}{|l|l|r|r|r|r|r|r|r|r|r|r|r|r|}
  \hline
	Strategy & Mutants & TinyXML2 & JSON & Google Test & CppCheck & Saab Case \\\hline
	
	\multicolumn{4}{l}{\vspace{-4mm}} \\
	\multicolumn{4}{l}{Generate Mutants \& validation:} \\\hline
	All &  & 0.09s & 0.64s & 3.79s & 2m 47.15s & Confidential \\\hline

	\multicolumn{4}{l}{Detect (Un)Reachable Mutants:} \\\hline
	Unoptimised & Reachable & \multirow{4}{*}{0.11s} & \multirow{4}{*}{14m 12.09s} & \multirow{4}{*}{7m 11.39s} & \multirow{4}{*}{7m 51.60s} & \multirow{4}{*}{2.63s}\\\cline{1-1}
	Schemata & /test suite &  &  & &  & \\\cline{1-1}\cline{1-2}
	Reachable & \multirow{2}{*}{Reachable} &  &  & &  & \\
	Schemata & \multirow{2}{*}{/test case} &  &  & &  & \\\cline{1-1}\cline{3-7}
	Split-stream &  & 0s & 0s & 0s & 0s & 0s\\\hline

	\multicolumn{4}{l}{\vspace{-4mm}} \\
	\multicolumn{4}{l}{Compile Mutants:} \\\hline	
	\multirow{3}{*}{Unoptimised}
		&  Reachable & 7m 17.49s & 21h 45m 46.62s & 40h 21m 45.49s & 3d 18h 52m 39.52s & \multirow{3}{*}{Confidential}  \\\cline{2-6} 
		& Unreachable & 22.79s & 0s & 2h 19m 35.25s & 10h 36m 20.31s &  \\\cline{2-6}  
		& Invalid & 50.83s & 3h 05m 23.74s & 3h 05m 24.35s & 1h 01m 41.15s &  \\\hline 
	Schemata & \multirow{4}{*}{Valid} & \multirow{3}{*}{1.23s} & \multirow{3}{*}{3m 54.68s} & \multirow{3}{*}{6m 36.04s} & \multirow{3}{*}{7m 07.39s} &  \multirow{3}{*}{3m 02.63s}\\\cline{1-1}
	Reachable &&&&&& \\
	Schemata &&&&&& \\\cline{1-1}\cline{3-7}
	Split-stream & & 1.28s & 3m 57.27s & 6m 41.85s  & 7m 13.91s & N.A.
\\\hline
	
	\multicolumn{4}{l}{\vspace{-4mm}} \\
	\multicolumn{4}{l}{Execute Mutants (excluding timed out):} \\\hline
	\multirow{2}{*}{Unoptimised}
		& Reachable & 1m 05.60s & 17h 41m 49.05s & 9h 49m 36.26s & 2d 05h 41m 30.47s &  \multirow{6}{*}{Confidential}\\\cline{2-6}
		& Unreachable & 4.10s & 0s & 36m 01.25s & 1d 03h 02m 43.46s &  \\\cline{1-6}
	\multirow{2}{*}{Schemata}
		& Reachable & 1m 21.84s & 17h 43m 28.57s & 9h 50m 30.45s &  4d 22h 16m 45.61s & \\\cline{2-6}
		& Unreachable & 4.81s & 0s & 36m 04.84s &  2d 11h 32m 05.66s & \\\cline{1-6}
	Reachable & \multirow{2}{*}{Reachable} & \multirow{2}{*}{1m 13.54s} & \multirow{2}{*}{1h 01m 29.01s} & \multirow{2}{*}{1h 46m 36.16s} &  \multirow{2}{*}{9h 27m 30.01s} & \\
	Schemata & \multirow{2}{*}{/test case} &  &  &  &  & \\\cline{1-1}\cline{3-7}
	Split-stream & & \multicolumn{5}{c|}{DNF}\\\hline
	
	\multicolumn{4}{l}{\vspace{-4mm}} \\
	\multicolumn{4}{l}{Technique Execution Overhead:} \\\hline
	Schemata vs
		& Reachable & 17.44\% & 0.16\% & 0.15\% &  120.29\% & \multirow{2}{*}{0.11\%} \\\cline{2-6}
	Unoptimised
		& Unreachable & 19.95\% & N.A.
& 0.17\% & 120.13\% & \\\hline

	\multicolumn{4}{l}{\vspace{-4mm}} \\
	\multicolumn{4}{l}{Timed Out Mutants:} \\\hline
	Unoptimised && \multirow{4}{*}{27.60s} & \multirow{4}{*}{0s} & \multirow{4}{*}{26m 00s} &  2d 12h 30m 25s & \multirow{4}{*}{Confidential}\\\cline{1-1}\cline{6-6}
	Schemata &&  &&&  \multirow{2}{*}{5d 01h 00m 50s} &\\\cline{1-1}
	Reachable &&  &&&  &\\
	Schemata &&  &&&  &\\\cline{1-1}\cline{3-7}
	Split-stream & & \multicolumn{5}{c|}{DNF}\\\hline
\end{tabular}
\end{table*}

\subsection{Individual Timings}

\myparagraph{Generate Mutants.}
The first block of \tabref{table:IndividualTimings} confirms that the generate mutants phase is negligible. This phase is necessary and identical for all our optimisation techniques and corresponds to $t_{\textit{mutant\_generation}}$ in Formula \ref{eq:unoptimised} to 4.
The four open-source cases illustrate that this phase is orders of magnitude faster than any of the following steps and less than the original compilation time of the projects as listed in \tabref{table:ProjectDetails}.

\myparagraph{Detect (Un)Reachable Mutants.}
In the second block of \tabref{table:IndividualTimings}, we listed the detection times of the (un)reachable mutants.
For the unoptimised and schemata technique, we detected which mutants were reached by the test suite.
The \hybrid approach is more fine-grained and detects which mutants are reached on a test-by-test case.
We estimated this phase at $\left(t^{\textit{schemata}}_{\textit{compilation}} + t^{\textit{schemata}}_{\textit{test\_suite\_execution}}\right)$, a single compilation and test suite execution in Formula \ref{eq:reachable schemata}.
The four open-source cases and the industrial case illustrate that this phase is also orders of magnitude faster than any of the following steps.

As the split-stream approach inherently only executes reachable mutants, it does not have a detection delay.

\myparagraph{Compile Mutants.}
\label{sec:Mutant Compilation}
The third block of \tabref{table:IndividualTimings} confirms that the compilation phase takes up a significant amount of time.
Additionally, the compilation of the invalid and unreachable mutants is considerable.

We estimated this phase for the unoptimised approach at $t_{\textit{compilation}}*(\textit{reachable\_mutants} + \textit{unreachable\_mutants} + \textit{invalid\_mutants})$ in Formula \ref{eq:unoptimised}.

By moving from the unoptimised approach to the schemata approach, we only need to compile once instead of for all mutants, reducing the compilation time to $t^{\textit{schemata}}_{\textit{compilation}}$ from Formula \ref{eq:schemata} where the multiplication is removed.
This essentially removes the compilation overhead as is confirmed by the four open source cases in \tabref{table:IndividualTimings}.
We can also see that the compilation is only slightly longer than the original compilation, as seen in \tabref{table:ProjectDetails}.
This is due to the fact that the injected mutants increase the code base, but this is limited to each function scope, therefore leaving the linking part of the compilation untouched.
However, when we introduce more advanced mutation operators and mutate the aforementioned const expressions, we likely need to introduce different functions, causing the compilation time to increase further.
Still, we can expect the compilation time to remain drastically decreased from an unoptimised traditional approach.
This is also the case for the split-stream mutation testing approach, which in its turn is slightly longer compared to the mutant schemata one, as it has internal functions to regulate the activation and forking of the mutants.

\myparagraph{Execute Mutants.}
The fourth block of \tabref{table:IndividualTimings} contains the execute mutant phase excluding the timed out mutants.
Here we can see that the execute mutant phase also takes up a significant amount of time.
This was expected as we estimated its duration at $t_{\textit{test\_suite\_execution}}* (\textit{reachable\_mutants} + \textit{unreachable\_mutants})$ for the unoptimised and schemata approach in Formula \ref{eq:unoptimised} and \ref{eq:schemata}.
The first observation we can make is that the unreachable mutants can have a minimal to large impact on the mutant test execution.
The stronger a test suite is, the fewer mutants will be completely unreachable.
We see no impact in the JSON project as it has no unreachable mutants.
We see the most impact in the CppCheck project.
Here, the test execution time of the unreachable mutants for the \textit{unoptimised and mutant schemata} approach is half of the reachable ones.
It is possible that running the project with a different configuration leads to a different number of unreachable mutants.

Our second observation is that the execution time of the \hybrid approach is drastically shorter than that of the schemata and unoptimised approach, even when compared to only the reachable mutants.
Instead of executing the entire test suite for each mutant, the \hybrid approach only executes those test cases that reach the specified mutant.
A mutant that survives will have fewer test cases executed compared to the mutant schemata and unoptimised approach.
It also stands to reason that mutants that are detected are detected faster as the test cases that do not reach the mutant are not executed.
We estimated its duration at $t^{\textit{schemata}}_{\textit{test\_suite\_execution}}*\textit{reachable\_mutants} * {\textit{decoupling\_factor}}$ in Formula \ref{eq:reachable schemata}.
In \tabref{table:IndividualTimings} we can see that the decoupling factor varies from project to project.
This factor can roughly be represented by the average amount of tests that will be executed per mutant, which we listed in \tabref{table:ProjectDetails}.
TinyXML2 has only 1 test, so there is no additional speedup to be gained from the \hybrid approach. 
The JSON project has the most speedup from this approach as it only needs to execute 9.61\% of the test cases per mutant, this is followed by the CppCheck project at 10.76\% and the Google Test project at 19.53\%. 

Our last observation is that we, unfortunately, could not gather data regarding the test suite execution of split-stream mutation testing on the projects under analysis.
We, therefore, labeled its result as DNF.
While we build in support to reset local files to a specific state using a local GitHub repository, and built in hook functions to reset more advanced dependencies like running, or even off-site, databases, we could not get this to work reliably for the tested projects.
Our projects all relied on open IO-streams which we cannot reset unless we reimplement or overload these functions.
For our projects, this meant that eventually, the IO files got corrupted and/or the file offset pointer.
We did not search for a solution as while this strategy offers an expected speedup of 2x, the strategy is difficult to incorporate with external dependencies such as databases as it requires in-depth knowledge of the system.
We, therefore, do not recommend using the strategy, at least not as a first option.

\myparagraph{Technique Execution Overhead.}
\label{sec:overhead}
In block four we can see that the execution time for both the unreachable and the reachable mutants is increased by using the schemata technique compared to the unoptimised technique.
This is due to the additional run-time instructions that ensure the correct activation of the mutants.
We listed the execution overhead introduced by the optimisation techniques in the execute mutant phase in percentages in the fifth block of \tabref{table:IndividualTimings}.
We can see that the percentual overhead between the unreachable and reachable mutants is, as expected, very close.
As there are no unreachable mutants for the JSON project, no overhead can be calculated for it.
We can also see that there is a limited overhead for the reachable mutants in the JSON and Google Test projects of 0.16 and 0.15\%.
The TinyXML2 project is impacted more by 17.44\%.
This increase in overhead can be attributed to the increase in the number of mutants per LOPC.
Where this was only 0.047 and 0.084 for JSON and Google Test, it is 0.241 for TinyXML2.
The more mutants there are per LOPC, the more if statements there are to verify which mutant needs to be executed.
Mathematically dense programs will suffer more from the schemata implementation using the ternary operator.
Switching to a switch-case implementation would reduce the impact.
The CppCheck case has, with 0.565 mutants per LOPC, the highest number of mutants per LOPC.
We thus also expected a higher impact on the execution time of the project.
We measured an overhead of 120.29\%.
This overhead can be attributed to the many if statements that are introduced per statement.
Changing the activation of the mutants from if statements to switch-case-based should reduce this overhead.

\myparagraph{Timed Out Mutants.}
For many of the projects, the test time caused by timed-out mutants is fairly limited.
This can be seen in the last block of \tabref{table:IndividualTimings}.
However, for the bigger, and longer running projects, this does become a significant amount of the mutation analysis time.
This is especially true for the CppCheck project as the timed-out mutants take up 93\% of the reachable schemata mutation analysis.
We manually set a fixed timeout time for each of the projects.
The time out is identical for each of the optimisation approaches except for the CppCheck project.
The introduced overhead caused an additional delay for the schemata-based approaches.
Here we double the time-out time.
The time-out time can be reduced specifically for the reachable mutant schemata strategy.
Here we execute each test case individually instead of the entire test suite.
We should thus set an individual time out for each of the test cases.
These time-outs will thus be much shorter, resulting in a reduced detection time when a mutant times out due to, e.g., an infinite loop.

\subsection{Complete Mutation Analysis}
In this section, we look at the impact of the optimisation strategies in relation to the complete mutation analysis whilst answering the research questions.
Their timings and speedups can be found in \tabref{table:speedups}.

\begin{table*}[hbt]
\caption{\protect{Results: speedups}}
\label{table:speedups}
\begin{normalsize}
\begin{tabular}{|l|r|r|r|r|r|r|r|r|r|r|r|r|}
  \hline
	  Mutants & TinyXML2 & JSON & Google Test & CppCheck & Saab Case \\\hline

	\multicolumn{4}{l}{\vspace{-4mm}} \\
	\multicolumn{4}{l}{Unoptimised:} \\\hline	
	\multirow{1}{*}{Considered}
		& 10m 12.42s & 1d 18h 33m 00.05s & 2d 09h 14m 27.64s & 11d 06h 50m 50.51s &  \multirow{5}{*}{Confidential} \\\cline{1-5}
	Valid 
		& 9m 17.58s & 1d 15h 27m 36.31s & 2d 05h 33m 02.04s & 10d 02h 46m 25.90s   &  \\\cline{2-5}
	(vs considered)
		& 1.10x & 1.08x & 1.07x & 1.12x   & \\\cline{1-5}
	Reachable
			& 8m 52.25s & 1d 15h 45m 50.43s & 2d 02h 44m 36.93s & 8d 13h 15m 13.74s   &  \\\cline{2-5}
	(vs valid)
	  		& 1.05x & 0.99x & 1.06x & 1.18x   & \\\hline

	\multicolumn{4}{l}{\vspace{-4mm}} \\
	\multicolumn{4}{l}{Schemata:} \\\hline	
	Valid
		& 1m 50.77s & 17h 47m 23.89s & 10h 59m 15.12s & 12d 10h 59m 35.81s   & \multirow{3}{*}{Confidential} \\\cline{2-5}
	(vs unopt.
valid)
		& 5.03x & 2.22x & 4.87x & 0.81x   &  \\\cline{1-5}
	Reachable
		& 1m 47.43s & 18h 05m 38.01s & 10h 30m 21.67s & 9d 23h 59m 35.81s   &  \\\cline{2-6}
	(vs unopt.
reachable)
		& 4.95x & 2.20x & 4.83x & 0.86x  & 5.16x\\\cline{2-6}
	(vs schemata valid)
		& 1.03x & 0.98x & 1.05x & 1.25x   & Confidential \\\hline
		
	\multicolumn{4}{l}{\vspace{-4mm}} \\
	\multicolumn{4}{l}{Reachable Schemata:} \\\hline	
	Reachable/test
		& 1m 43.92s & 1h 23m 38.45s & 2h 26m 27.38s & 5d 10h 46m 06.16s   & \multirow{5}{*}{Confidential} \\\cline{2-5}
	(vs unopt.
considered)
	 	& 5.89x & 30.52x & 23.45x & 2.07x   & \\\cline{2-5}
	(vs unopt.
reachable)
	 	& 5.12x & 28.52x & 20.79x & 1.57x   & \\\cline{2-5}
	(vs schemata reachable)
	 	& 1.03x & 12.98x & 4.30x & 1.83x   & \\\cline{2-5}	 	
	(vs valid schemata)
	 	& 1.07x & 12.76x & 4.50x & 2.29x   & \\\hline	 	
\end{tabular}
\end{normalsize}
\end{table*}

\subsubsection{RQ1: \RQOne}
The \textit{schemata approach}, which virtually eliminates the compilation overhead, drastically speeds up the mutation analysis in most cases.
As we have seen in \secref{sec:Mutant Compilation}, mutant schemata can reduce the compilation time down to almost the original, single, compilation time of the project.
Here we see speedups of up to 5.03x for the valid mutants, and up to 4.95x for the reachable mutants compared to the unoptimised approach, as listed in \tabref{table:speedups}.
JSON has a speedup of 2.22x due to a high proportion of test suite execution time versus its compilation time.
The CppCheck project obtained a speedup 0.81x due to the increase in mutant execution time, attributed to the techniques implementation in combination with the high number of mutants/LOPC.
In the previous section we saw that there was a limited impact on the execution times of the mutants in projects with a low number of mutants per line of production code (less than 0.1: JSON and Google Test), projects with a high number of mutants per line of production code (0.24 and 0.56: TinyXML2 and CppCheck) suffered from increased execution overheads which limited its speedup potential.
We proposed mitigating the overhead issue by using \emph{switch} statements instead of \emph{if} statements for mutant selection.
However, we leave the implementation and evaluation of this mitigation as future work.
But, we then expect to also see a speedup for this project.

We measured a speedup of 5.16x for the industrial Saab project for the mutant schemata approach compared to a traditional approach when excluding the completely unreachable mutants.
For the other projects we see that the speedups for this optimisation are slightly smaller than with the unreachable mutants.
This is as expected as more time (compilation and execution time) is saved excluding the unreachable mutants in an unoptimised approach then in a schemata approach (execution time).

\hypobox{
With the TinyXML2, Google Test and the industrial Saab project, the mutant schemata technique obtains a speedup between 4.87x and 5.16x.
For the other cases the speedup was less eminent.
JSON has a speedup of 2.22x due to a high proportion of test suite execution time versus its compilation time.
CppCheck obtained as speedup of 0.81x  due to due to the increase in mutant execution time, attributed to the techniques implementation in combination with a high number of mutants/LOPC.

The obtained speedup from a mutant schemata technique depends on two factors: (a) the proportion of the project compilation time versus its execution time; (b) the amount of mutants per line of production code.
The latter is implementation specific and its impact can be reduced by changing the activation of the mutants from if statements to switch-cases.
}

\subsubsection{RQ2: \RQTwo}
For \textit{split-stream mutation testing} we expected an additional speedup of approximately 2x over mutant schemata.
Unfortunately, we could not gather data regarding the test suite execution on the projects under analysis due to IO dependent issues.
While we build in support to reset local files to a specific state using a local GitHub repository, and built in hook functions to reset more advanced dependencies like running, or even off-site, databases, we cannot recommend this optimisation technique as it cannot easily be incorporate with external dependencies such as databases as it requires in-depth knowledge of the system.

\hypobox{
In theory this strategy could yield a speedup of a factor 2, but in practice applying the strategy proves to be too demanding.
Reverting to the internal state demands too much specific knowledge about the design of the system under test, especially in the case data is stored in external databases and filesystems.
}

\subsubsection{RQ3: \RQThree}
In \tabref{table:speedups} we see that for the \textit{unoptimised approach} a speedup between 1.07x to 1.12x is achieved by excluding the invalid mutants.
This speedup is attained by a reduction in the compilation time as invalid mutants are mutants that cause compilation issues and can thus not be executed.

Compiler-integrated techniques like mutant schemata and split-stream for the C language family come with an important drawback: the tight integration with the compiler.
Since all mutants are injected simultaneously, the resulting program must compile without any errors.
Invalid mutants are not acceptable, since they prevent the compilation (and the execution) of the mutated program.
This is especially challenging for statically typed languages with many interacting features and unforgiving compilers (C, C++, \ldots).
Since \Clang has access to all of the project's type-information, we can programmatically ensure that any of the created mutants are statically correct.
We can therefore not calculate a speedup for the invalid mutants as their exclusion is a requirement for the schemata techniques.

\hypobox{
Compared to an \textit{unoptimised} approach the reduction in the compilation overhead leads to a speedup by a factor between 1.07x and 1.12x.
This is not all that much, but excluding invalid mutants is a necessary prerequisite for the mutant schemata strategy discussed under RQ1.}

\subsubsection{RQ4: \RQFour}
In \tabref{table:speedups} we see that for the \textit{unoptimised approach} a speedup between 1.05x to 1.18x is achieved by excluding the completely unreachable mutants, except for the JSON project as it has no unreachable mutants
This speedup stays approximately the same with the \textit{mutant schemata} approach where the speedup is between 1.03x to 1.25x by excluding the completely unreachable mutants, except for the JSON project as it has no unreachable mutants.
This speedup, however, will depend on the coverage of the test suite. A test suite with low coverage, and thus reaching fewer mutants will yield a higher speedup by excluding them.

However, the \textit{reachable schemata} technique goes a step further by not only excluding completely unreachable mutants but by also excluding the test cases which do not reach the mutant on a mutant by mutant case.
This gives an additional speedup between 1.83x and 12.98x over the schemata technique that excludes completely unreachable mutants.
This provides a speedup between 2.29x and 12.76x over the normal schemata technique, except for the TinyXML2 project, as it only has a single test case.
Further speedups are possible by optimising the time-out function that e.g. detects mutants stuck in infinite loops.
This can be achieved by specifying a time-out for each test case instead of a time-out for the global test suite.

\hypobox{
Compared to an \textit{unoptimised} approach, excluding the completely unreachable mutants provides a speedup between 1.05x to 1.18x.
If we go one step further (also excluding the test cases which do not reach the mutant on a mutant by mutant base) we achieve an additional speedup between 1.83x and 12.98x.
}

\section{Limitations and Lessons Learned}
\label{sec::LesonsLearned}
In this section, we will derive the limitations and lessons learned geared toward the mutation testing community.

Mutant schemata for C++ can bring a performance improvement of an order of magnitude by eliminating the compilation overhead.
However, mutant schemata for C++, and all strongly typed languages, require the ability to extract extra information to ensure that none of the mutants causes compilation errors as all mutants need to be compiled at once.
We achieved this by using \Clang, allowing us to extract the statically available information and to create the mutants in the actual source code.
Mutant schemata and the additional speedup strategies also require instrumentation of the source code.
For this, we also relied on \Clang.
Doing our work we learned 10 lessons, which we list below.

\myparagraph{Type Safety.}
Some challenges occur when we try to implement mutant schemata for statically typed programming languages like C++.
First and foremost, the mutated program must be syntactically correct and no type errors should occur.
This means that every mutated statement should be valid.
We cannot generate mutants like ``string - string'' or ``float \% int''.
Classes can implement or omit operators like `+' and `-' further complicating the matter.
 \Clang allows us to access all the statically available information of the project and to verify if a mutated statement is syntactically correct without the need to compile the complete project.

While we currently have this working for binary expressions, we have not investigated how this needs to be done for less straightforward mutations such as Access Modifiers Change (AMC), where e.g. the public access label is changed to private.
While we believe this can be done using \Clang, we envision that the analysis time for such a change will be longer compared to the analysis of a binary expression.

\myparagraph{Ternary Operator.}
Using the ternary operator also somewhat limits the mutation kinds we can use, as all mutants in the expression need to be of the same type.
We cannot have a char pointer on the \textit{iftrue} side and en unsigned int on the \textit{iffalse} side.
Fortunately, virtually all mutations on a single expression will have the same type or can be dynamically casted to that type.

\myparagraph{Mutation Operator Support.}
Currently, for our proof-of-concept tool, we have support for Relational Operator Replacement (ROR), Arithmetic Operator Replacement (AOR) and Logical Connector Replacement (LCR).
Our proof-of-concept tool can be easily extended to support other binary mutation operators and unary operators.
However, other mutation operators, like the previously mentioned Access Modifiers Change (AMC) might need more work.

\myparagraph{Const, constexpr, and templates.}
The driver for mutant schemata relies on information from outside the program to control the activation of the mutants by setting the MUTANT\_NR variable.
The MUTANT\_NR variable is initialised at runtime and will thus never be const.
This means that we cannot use the variable inside const and constexpr functions, as these functions are evaluated at compile time and the MUTANT\_NR value cannot be known at compile time.
An example of this can be seen in Listing \ref{lst:Invalid schemata from const}.

\begin{fancylisting}{Invalid Schemata from const,label=lst:Invalid schemata from const}
const float a = 1.2;
const float b = 2.0;
// original statement
const float r = a * b;
 
// mutated statement
extern int MNR;		// driver for active mutant
const float r = (MNR == 1 ? a + b :
				(MNR == 2 ? a - b :
				(MNR == 3 ? a / b : 1.2 * 2)));
						   (*@ \textasciitilde\textasciitilde\textasciitilde \textasciicircum\textasciitilde\textasciitilde\textasciitilde \textcolor{red}{Invalid schemata MNR cannot be know at compile time} @*)

\end{fancylisting}

This means that we cannot mutate const and constexpr in the same way as non-const functions.
This includes type definitions (e.g. \textit{using ...}), template arguments, static\_asserts, etc.

For now, we choose not to implement support for these kinds of mutations.
We do however generate and verify these for a standalone mutation, but using them in a single compilation for mutation testing using mutant schemata and/or split stream requires additional logic.
This can be implemented by creating separate values and/or functions for each mutated operation and selecting the correct one everywhere in the project where they are used (e.g. const val becomes const val\_0, const val\_1, const val\_2, ...).
This will drastically increase the size of the code base and the compiled binary.
Additionally, this will reduce the readability of the code, however, normal developers should not see this.

\myparagraph{Switch-Case.}
In \secref{sec:Schemata Implementation} we explained the advantages of using the ternary operator for a mutant schemata approach.
We envisioned that there would be a fixed overhead cost for the technique per mutant, but that this overhead would remain limited.
In our results we saw that this was only the case for projects with a low to medium number of mutants per LOC.
For projects where the number of mutants per line of production code is high, such as math-heavy projects, the ternary implementation causes too many additional evaluations of the if statements to reach the correct mutant.
The most common case where no mutant in the statement is activated is the worst-case scenario, as all the if statements need to be evaluated before the original statement can be executed.
In the simplistic example in Listing \ref{lst:switch}, we can see that an ``$a*b*c$'' statement has six mutations.
This means that for every mutant outside of that expression, six if statements would need to be evaluated.
By switching to a switch-case implementation like in Listing \ref{lst:switch}, only the switch needs to be evaluated and the offset for which case to jump to needs to be calculated.
This would save many instructions especially when there are many instructions on a single line of code.
This would have greatly benefited performance in the CPPCheck case.

To reduce the overhead of the schemata strategy, we recommend using the switch-case implementation in larger projects, despite the implementation challenges that need to be overcome.
We described some of these challenges and tradeoffs in \secref{sec:Schemata Implementation}, including code explosion and the scoped nature of the switch case causing the need to initialise variables outside the switch.
Because of these challenges, specific research will need to be performed to understand where and how switch cases should be used, and how to deal with its tradeoffs.
We therefore deem this as out-of-scope for the current research, but as something that needs to be investigated in future work.

\begin{fancylisting}{Mutant Schemata using Switch Case,label=lst:switch}
extern int MNR;		// driver for active mutant
float f(float a, float b, float c) {
    switch (MNR) {
        case 1:  return (a * b) + c;
        case 2:  return (a * b) - c;
        case 3:  return (a * b) / c;
        case 4:  return a + (b * c);
        case 5:  return a - (b * c);
        case 6:  return a / (b * c);
        default: return a * b * c;
    }
}
\end{fancylisting}

\myparagraph{Test per test case vs per module.}
Our program extracts the reachable mutants per test case or per module for the program.
The best results are obtained by using fine-grained test selection and running each test case individually instead of module-grained.
All our projects are compatible with the fine-grained test selection.
Testing per test instead of per module reduces the AVG number of tests per mutant.
We verified this for the CPPCheck project. Here the AVG tests/mutant for the fine-grained test selection is 403, while it is 812 for the module-based one.
The CPPCheck program implemented its own testrunner to enable running the tests per test instead of per module.
If you use ctest, then traditionally, tests are added using the \textit{add\_test} command to a specific test module.
Each test module can be run separately, but you cannot run a specific test within a test module without running the other tests from that module.
By switching from the \textit{add\_test} command to \textit{gtest\_add\_tests} and/or \textit{gtest\_discover\_tests}, one can run each test individually.
This means that each test will run slightly slower, as a new test environment will be created for each test instead of for each module.
This will however be greatly offset by the reduction in the amount of mutants that will be selected to run for each test.

\myparagraph{Multi-threading.}
Traditional, unoptimised mutation testing and mutant schemata inherently support applications that utilise multiple threads as they run from start to finish.
Split-stream mutation testing on the other hand does not.
If we `start' the first mutant after a secondary thread was spawned, our mutant will likely run to the end of the program and wait for the secondary thread to finish.
A second mutant `started' from the same location will expect the secondary thread to be there, but as it is already closed, the second mutant will not run correctly.
Supporting multi-threaded programs for the split-stream approach will require additional development to ensure that not only the main thread will be forked, but also the existing threads.

\myparagraph{External Dependencies.}
Traditional, unoptimised mutation testing and mutant schemata are highly likely to support external dependencies.
This is due to the fact that external dependencies are a common phenomenon in Continuous Integration settings.
After the test suite has run, the external dependencies, like input/output files and databases, need to be reset for the next run of the Continuous Integration.
The support for this is usually baked into the used build systems of the projects (e.g. make clean).

Split-stream mutation testing on the other hand requires the external dependencies to be able to be reset to a specific state.
While we build in support to reset local files to a specific state using a local GitHub repository, more advanced dependencies like running, or even off-site, databases will need very specific commands and domain knowledge to enable the strategy to reset them to a specific state.
The implementation for this will be different for each project.

These problems are not insurmountable, they are similar in nature to the problems that occur with distributed mutation testing.

\myparagraph{Split-Stream Mutation Testing.}
Split-stream mutation testing allows us to reduce the execution overhead by an approximate factor of 2.
However, this strategy requires too much knowledge of the system.
It cannot easily be incorporated with external dependencies, e.g. databases.
Our current recommendation is to not use the strategy.

\myparagraph{Timed Out Mutants Overhead.}
For the bigger projects like CppCheck, we have seen that the execution time of the optimised mutation analysis consists mostly of timed-out mutants. Here, the timed-out mutants take up 93\% of the reachable schemata mutation analysis.
In our current approach, we detected the timeout once the total time for that mutant reached a threshold.
For the CppCheck project, this means that a test that is stuck in an infinite loop would only be timed out after 45 seconds.
As the CppCheck project has 3,745 test cases, the average test time is below 0.01 seconds.
The impact of the timed-out mutants will be drastically reduced if we stop mutants not after a global threshold but after localised thresholds based on the individual tests.

%%%%%%%%%%%%%%%%%%%%%%%%%%%%%%%%%%%

\section{Threats to Validity}
\label{sec::Threats}
As with all empirical research, we identify those factors that may jeopardise the validity of our results and the actions we took to reduce or alleviate the risk.
Consistent with guidelines for empirical research (see~\cite{case_study_research_SWE,case_study_research}), we organise them into four categories.

\myparagraph{Construct validity: } do we measure what was intended?\\
In essence, we want to know which parts of the mutation testing process are affected by building a mutant schemata and split-stream mutation optimisation on top of the \Clang front-end, as we believed these strategies would eliminate the compilation and execution overhead of mutation testing.
The mutant schemata strategy and extended reachable schemata strategy cause additional delays in some parts of the mutation testing process, but in return eliminate the need to compile every mutant individually.
The real question is whether the time benefit from a single compilation outweighs the added delays.
For this, we measured each part of an unoptimised mutation testing process and compared it to the optimisation strategies.

We implemented all of our mutants using the ternary (conditional) operator.
This causes many additions to the code base, which impact the compilation time and test suite execution time.
As mentioned before, this caused a large delay in the test suite execution time for the CheckCpp project as many ternary operators are written on frequently executed lines.
In such cases, a different structure using switch-cases could have prevented such large delays.

While we chose a limited set of mutant operators (i.e. ROR, AOR, and LCR) and omitted mutants in \emph{const} and templated expressions, we believe that our results represent the performance benefits one can achieve using the strategy.
We believe so because the compilation times for the currently supported mutants are only impacted slightly, adding less than 25\% to the compilation time for our biggest project.
The test suite execution times were not impacted significantly except for the CheckCpp case where it caused an additional delay of 120\%, which could be improved by using \emph{switch} statements instead of \emph{if} statements to select the mutants.

To ensure that the results are reliable and comparable, we used the same generate mutants method for all optimisation techniques. Additionally, we verified that the mutants we label as invalid mutants using the \Clang front-end are the same mutants that actually caused compilation failures in the unoptimised baseline. This ensures that the same mutants are used across the various steps of the optimisation techniques.

\myparagraph{Internal validity:} are there unknown factors which might affect the outcome of the analyses?\\
We expected that the more mutants there are per line of code, the more the test suite execution time would be impacted as there is simply more code to execute.
However, most of our projects did not show such results, they showed no difference in execution time with and without mutants.

As mentioned before, we believe that this might be due to the architecture of modern CPUs where out-of-order execution and pipeline depths/stalls can influence the effective performance.
However, we did not investigate this sufficiently to draw firm conclusions.

We, however, did verify this in the case of CppCheck, for which the test execution times were heavily impacted.
This was due to the number of mutants that can occur on a single line, drastically changing the execution time of that specific line.
When that line is executed frequently, the performance of the entire system is affected considerably.
On the other hand, the performance impact is minor if that single line is only executed once.
Ideally, we need to investigate this further using a profiler.

To minimise environment factors, we locked the frequency of the CPU to the base frequency of the CPU and kept the computer in a well-ventilated space to prevent CPU throttling.
While we used an older computer (Intel(R) Core(TM)2 Quad Q9650 CPU) architecture to run our experiments, we believe that this does not impact the analysis.
The use of a hard drive instead of an SSD however will reduce the compilation times and likely lower the total speedups.
We believe, however, that this should be limited, and that the benefits of the optimisation techniques would still be valid even when using an SSD.
 
We used the same infrastructure for the analysis of the five selected open-source projects.
We used an Intel(R) Core(TM)2 Quad Q9650 CPU, with two 4GB (Samsung M378B5273DH0-CH9) DDR3 RAM modules (for a total of 8GB) and a 250GB Western Digital (WDC WD2500AAKX-7) hard drive.
The PC was running Ubuntu 18.04.1 LTS (GNU/Linux 4.15.0-29-generic x86\_64).
Using an SSD would drastically influence the compilation times and negatively impact the total speedups.
Using more RAM and/or a faster CPU might influence the compilation time and/or execution time, this, however, is presumed to be marginal.

\myparagraph{External validity:} to what extent is it possible to generalise the findings?\\
We evaluated our proof-of-concept tool on four open-source projects with different characteristics and one industrial project.
These projects vary in size and in computational needs, varying from a low number of mutants per line of production code to high numbers.
We believe that our results from these different projects represent the performance benefits one can achieve by using the different optimisation strategies.

\myparagraph{Reliability:} is the result dependent on the tools?\\
We took great care to ensure that external tools did not impact our timing results.
The internal toolchain builds on very established components from the \LLVM and \Clang projects.
We only measured the time it took to compile the project, generate the mutants, and execute them.
We excluded any timings related to external tools or implementations like database access times to store the mutants.
Our results thus represent the performance benefits of the optimisation strategies as implemented.
On the other hand, we do see areas which can affect the performance.
In some cases, where many mutants are listed on a single line, it might be faster to execute them using a \emph{switch} statement than to use our \emph{if} statement.
In this case, the compiler can create a jump table (using consecutive indexes), i.e. an array of pointers, to directly jump to the correct label.
Adding more mutation operators might also impact the performance.

%%%%%%%%%%%%%%%%%%%%%%%%%%%%%%%%%%%

\section{Conclusion}
\label{sec::Conclusion}
In this paper we investigate to which extent the \Clang front-end and its state-of-the-art program analysis facilities allow to implement existing strategies for mutation optimisation within the C language family.
We present a proof-of-concept tool that allows us to collect detailed measurements for each of the mutation phases, i.e. generate mutants, compile mutants, and execute mutants.
We validate the proof-of-concept tool on four open-source C++ libraries and one industrial component, covering a wide diversity in size, C++ language features used, compilation times, and test execution time.
As such, we analyse the speedup from two perspectives (compilation time and execution time) assessing four optimisation strategies (exclude invalid mutants, mutant schemata, reachable mutant schemata, split-stream mutation testing).
We address four research questions.

\myparagraph{RQ1: \RQOne}
We could virtually eliminate the compilation overhead to the point that the compilation for all mutants was only slightly longer compared to the original compilation time, i.e. the time it takes to compile the project without any mutants.
With the TinyXML2, Google Test and the industrial Saab project, the mutant schemata technique obtains a speedup between 4.87x and 5.16x.
Yet, the obtained speedup depends on two factors: (a) the proportion of the project compilation time versus its execution time; (b) the amount of mutants per line of production code.
The latter is implementation specific and its impact can be reduced by changing the activation of the mutants from if statements to switch-cases.
As a consequence of our implementation, for the other cases the speedup was less eminent: JSON has a speedup of 2.22x due to a high proportion of test suite execution time versus its compilation time.
CppCheck obtained as speedup of 0.81x  due to the increase in mutant execution time, attributed to the techniques implementation in combination with a high number of mutants/LOPC.

\myparagraph{RQ2: \RQTwo}
In theory this strategy could yield a speedup of a factor 2, but in practice applying the strategy proves to be too demanding.
Indeed, the split-stream mutation technique demands that dependencies need to be revertible to specific states of the execution.
Reverting to the internal state demands too much specific knowledge about the design of the system under test, especially in the case data is stored in external databases and filesystems.
Thus for the cases we investigated, we could not eliminate the execution overhead with the split-stream mutation testing strategy.

\myparagraph{RQ3: \RQThree}
Compared to an \textit{unoptimised} approach the reduction in the compilation overhead leads to a speedup by a factor between 1.07x and 1.12x.
This is not all that much but it is a necessary prerequisite for the mutant schemata strategy.
Since all mutants are injected simultaneously, the resulting program must compile without any errors.
Invalid mutants are not acceptable, since they prevent the compilation (and the execution) of the mutated program.
This is especially challenging for statically typed languages with many interacting features and unforgiving compilers (C, C++, \ldots).
Since \Clang has access to all of the project's type-information, we can programmatically ensure that any of the created mutants are statically correct.

\myparagraph{RQ4: \RQFour}
Compared to an \textit{unoptimised} approach, excluding the completely unreachable mutants provides a speedup between 1.05x to 1.18x.
One notable exception is the JSON project which had no unreachable mutants.
This speedup stays approximately the same with the \textit{mutant schemata} approach where the speedup is between 1.03x to 1.25x.
This speedup, however, depends on the actual coverage of the test suite: a test suite with low coverage (thus reaching fewer mutants) yields a higher speedup.
The \textit{reachable schemata} technique goes one step further by not only excluding completely unreachable mutants but also excluding the test cases which do not reach the mutant on a mutant by mutant base.
This gives an additional speedup between 1.83x and 12.98x over the schemata technique that excludes completely unreachable mutants.
Providing a speedup between 2.29x and 12.76x over the normal schemata technique.
Further speedups are possible by optimising the time-out function that e.g. detects mutants stuck in infinite loops.
This can be achieved by specifying a time-out for each test case instead of a time-out for the global test suite.

\myparagraph{Overall.}
In summary, we successfully demonstrated the feasibility of using the \Clang compiler front-end for different optimisation strategies.
With the reachable schemata strategy, we virtually eliminated the compilation overhead to the point that the compilation for all mutants was only slightly longer compared to the original compilation time and we reduced the execution overhead by only executing the test cases for individual mutants which actually reach said mutants.
Compared to an \textit{unoptimised} approach we achieve a maximum speedup of 23.45x and 30.52x on the JSON and Google Test projects with the \textit{reachable schemata} strategy.
Even for less ideal scenarios from the CPPCheck and TinyXML2 projects we achieve a speedup of 2.07x and 5.89x.
These can be speedup further by two optimisations:
Firstly, use \emph{switch} statements for the mutant selection to reduce the technique overhead.
Secondly, tailoring the time-out function, that e.g. detects mutants stuck in infinite loops, on a test by test basis instead of on the global test suite.

Finally, we report some lessons learned for deploying a mutant schemata tool using the \Clang compiler framework.
Most important is that we need a different, specialised approach for generating mutants in \emph{const}, \emph{constexpr}, \emph{templates}, and \emph{define} macro's.
These statements are evaluated at compile-time, thus obstruct the runtime selection of the mutant required by the mutant schemata technique.

%%%%%%%%%%%%%%%%%%%%%%%%%%%%%%%%%%%

\section*{Acknowledgments}
This work is supported by
(a) the Research Foundation Flanders (FWO) under Grant number 1SA1519N; %% FWO grant Sten Vercammen
(a) the FWO-Vlaanderen and F.R.S.-FNRS via the Excellence of Science project 30446992 SECO-ASSIST.

%%%%%%%%%%%%%%%%%%%%%%%%%%%%%%%%%%%

%\backmatter

%\nocite{*}% Show all bib entries - both cited and uncited; comment this line to view only cited bib entries;
\bibliography{Vercammen2022STVRbib}%

\section*{Author Biography}

\begin{biography}{\includegraphics[width=66pt]{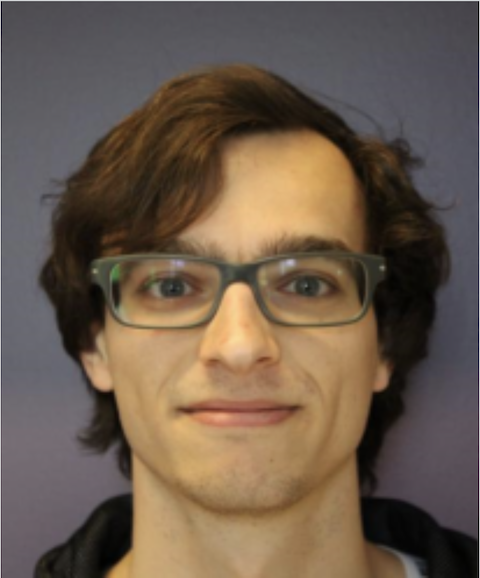}}{
\textbf{Sten Vercammen}
is a joint PhD student at the University of Antwerp (Belgium) and Lund University (Sweden).
Sten Vercammen is a member of the ANSYMO (Antwerp System Modelling) research group and under the supervision of Serge Demeyer, G\"{o}rel Hedin, and Markus Borg.
His main research is focused on speeding-up mutation testing for the C language family in order for it to become usable in an industrial setting.
To this end he has implemented a variety of optimisation techniques, from code instrumentation to compiler optimisations, and validated them on open-source projects and industrial projects.
}\end{biography}

\begin{biography}{\includegraphics[width=66pt]{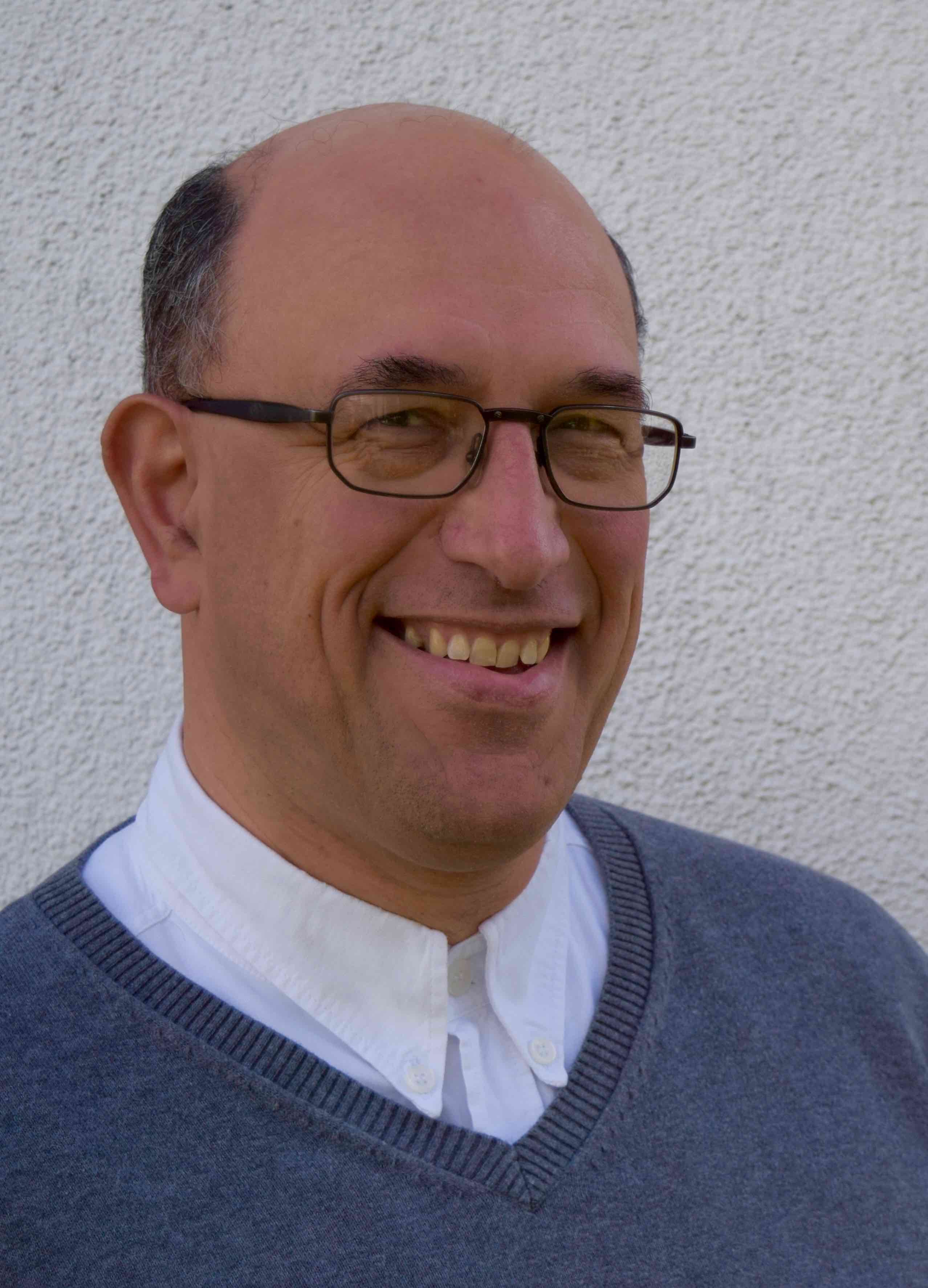}}{
\textbf{Serge Demeyer}
is a professor at the University of Antwerp and member of the ANSYMO (Antwerp System Modelling) research group.
Serge Demeyer is the chair of the NEXOR interdisciplinary research consortium and an affiliated member of the Flanders Make Research Centre.
In 2007 he received a ``Best Teachers Award" from the Faculty of Sciences at the University of Antwerp and is still very active in all matters related to teaching quality.
His main research interest concerns software evolution, more specifically how to strike the right balance between reliability (striving for perfection) and agility (optimising for adaptability).
He is an active member of the corresponding international research communities, serving in various conference organisation and program committees.
He has written a book entitled ``Object-Oriented Reengineering'' and edited a book on ``Software Evolution''.
He also authored numerous peer reviewed articles, many of them in top conferences and journals. He has an h-index of 42 according to Google Scholar.
Serge Demeyer completed his M.Sc. in 1987 and his PhD in 1996, both at the ``Vrije Universiteit Brussel''.
After his PhD, he worked for three years in Switzerland, where he served as a technical co-ordinator of a European research project.
Switzerland remains near and dear to his heart, witness the sabbatical leave during 2009-2010 at the University of Z\"urich in the research group SEAL.
The rest of Europe is explored territory as well, with a sabbatical stay in Lille, France (INRIA-RMOD) and in Lund, Sweden (RISE-SICS and SERG).
}\end{biography}

\begin{biography}{\includegraphics[width=66pt]{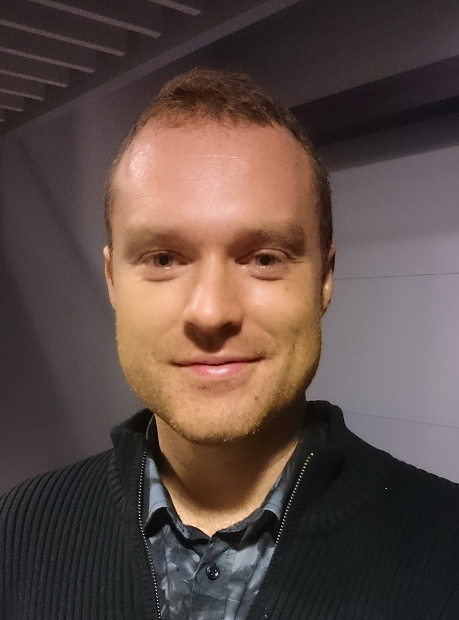}}{
\textbf{Markus Borg}
is a senior researcher within Mobility and Systems at RISE Research Institutes of Sweden. Furthermore, he is an adjunct lecturer at Lund University with the Dept. of Computer Science and a member of the Software Engineering Research Group (SERG). Markus Borg completed his MSc. in 2007 and his PhD in 2015, both at Lund University. After his MSc. degree, he worked three years in a safety-critical context at ABB with compiler and editor development for the process automation domain. Since 2015, he is primarily active in the  requirements engineering and software testing research communities, with a particular focus on AI engineering for the Swedish automotive industry. Markus Borg was a board member of Swedsoft between 2018-2022, a trade organisation acting as the voice of the Swedish software industry. He is now on the editorial board of the journal Empirical Software Engineering and acts as the department editor for the Requirements column in IEEE Software.
}\end{biography}

\begin{biography}{\includegraphics[width=66pt]{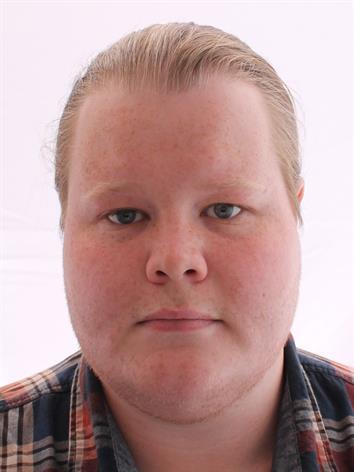}}{
\textbf{Niklas Pettersson}
Niklas Pettersson is a Software Engineer at Saab Aeronautics, Link\"{o}ping, Sweden, where he works at the Avionics Systems area (which is responsible for the development of the aircraft platform (computer architecture, software architecture, implementation of low-level and safety-critical software) as well as the development process and methodology in software development).
\vspace{1em}
}\end{biography}

\begin{biography}{\includegraphics[width=66pt]{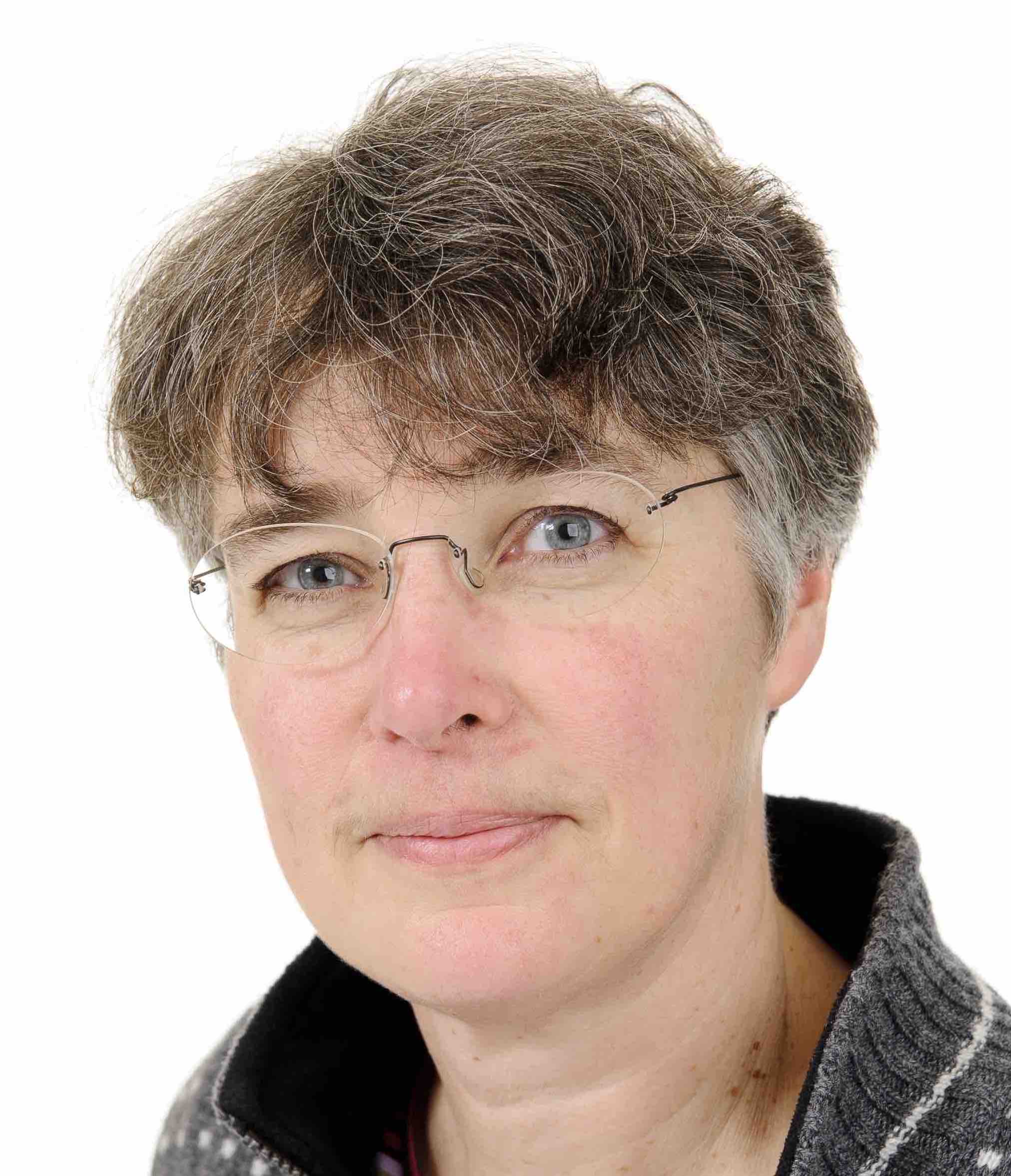}}{
\textbf{G\"{o}rel Hedin}
is a professor at Lund University, Sweden, where she heads the Software Development and Environments group.
Her research is focused on language tools like compilers, visual program editors, and source code analyzers.
She has developed reference attribute grammars as a technique for generating extensible language tools.
She leads the development of the open source metacompiler JastAdd that is the main tool for reference attribute grammars, and which is used in both academia and industry.
G\"{o}rel Hedin is active in the international programming language research community, and has served in numerous program committees for, e.g., OOPSLA, ECOOP, and SLE.
}\end{biography}

\subsection*{Author contributions}
Sten Vercammen conceptualised the paper, implemented the proof-of-concept tool, curated the experiment data, performed the analysis of the data, and wrote the paper. Niklas Pettersson ran the experiments on the Saab project and aided with the experimental integration into \Dextool. The remaining authors reviewed and edited the paper, whilst Serge Demeyer and Markus Borg also supervised the project.

\subsection*{Financial disclosure}

This work is supported financially by The Research Foundation -- Flanders (FWO), a public funding organisation in Belgium.
There is no direct or indirect industrial support for the research reported here.

\subsection*{Conflict of interest}

The authors declare no potential conflict of interests.

\section*{Supporting information}

The following supporting information is available as part of the online article:

\begin{table*}[hbt]
	\begin{tabulary}{\textwidth}{|L|p{0.6\textwidth}|}
	\hline
	\textbf{Availability of data} & \textbf{Availability statement} \\\hline
	\multirow{2}{=}{Data openly available in a public repository that does not issue DOIs} & The data that support the findings of this study are openly available in Github: \\
	&	\begin{minipage}[t]{0.6\linewidth}
		\begin{tabular}{ll}
			\parbox{\widthof{Google Test at}}{F-ASTMut at} & \url{https://github.com/Sten-Vercammen/F-ASTMut}
		\end{tabular}
		\end{minipage}  \\\hline
	Data derived from public domain resources & The data that support the findings are derived from the following resources available in the public domain: \\
	& 	\begin{minipage}[t]{0.6\linewidth}
		\begin{tabular}{ll}
			TinyXML2 at & \url{https://github.com/leethomason/tinyxml2} \\
			JSON at & \url{https://github.com/nlohmann/json} \\
			Google Test at & \url{https://github.com/google/googletest} \\
			CppCheck at & \url{https://cppcheck.sourceforge.io}
		\end{tabular}
		\end{minipage}  \\\hline

	Data subject to third party restrictions & The data that support the findings of this study are from Saab AB. Restrictions apply to the availability of these data. The available data are inside this article with permission of Saab AB.\\\hline	
	\end{tabulary}
\end{table*}

%\begin{table*}[hbt]
%	\begin{tabulary}{\textwidth}{|L|l|L}
%	\hline
%	\textbf{Availability of data} & \multicolumn{2}{L}\textbf{Template for data availability statement}} & \\\hline
%	\multirow{2}{=}{Data openly available in a public repository that does not issue DOIs} & \multicolumn{2}{L}{The data that support the findings of this study are openly available in Github:} & \\
%	& F-ASTMut at & \url{https://github.com/Sten-Vercammen/F-ASTMut} \\\hline
%	Data derived from public domain resources & \multicolumn{2}{L}{The data that support the findings are derived from the following resources available in the public domain:} & \\\hline
%	Data subject to third party restrictions & &\todo{} \\\hline	
%	\end{tabulary}
%\end{table*}

\end{document}